\documentclass[a4paper,oneside,onecolumn]{aa}
\usepackage{times}
\usepackage{graphics}

\makeatletter

\newcommand{\LyX}{L\kern-.1667em\lower.25em\hbox{Y}\kern-.125emX\@}
\newcommand{\noun}[1]{\textsc{#1}}
\let\SF@@footnote\footnote
\def\footnote{\ifx\protect\@typeset@protect
    \expandafter\SF@@footnote
  \else
    \expandafter\SF@gobble@opt
  \fi
}
\expandafter\def\csname SF@gobble@opt \endcsname{\@ifnextchar[
  \SF@gobble@twobracket
  \@gobble
}
\edef\SF@gobble@opt{\noexpand\protect
  \expandafter\noexpand\csname SF@gobble@opt \endcsname}
\def\SF@gobble@twobracket[#1]#2{}

\def\dvHartwick{2} 
\def\nGCutili{31}  
\def\reduced{52} 
\def\nDutchText{Thirty-nine}  
\def\nJKT{16} 
\def\mMUnMetro{16} 

\makeatother

\begin{document}

\thesaurus{05(08.01.1; 08.16.3 ; 11.01.1 ; 10.07.2)}

\title{The red giant branches of Galactic globular clusters in the \protect\( [(V-I)_{0},M_{V}]\protect \)
plane: metallicity indices and morphology }

\author{Ivo Saviane \inst{1} \and Alfred Rosenberg \inst{2,3} \and Giampaolo Piotto\inst{1} \and Antonio
Aparicio \inst{4}}

\offprints{Ivo Saviane}

\mail{saviane@pd.astro.it}

\institute{Dipartimento di Astronomia, Universit\`a di Padova, vicolo dell'Osservatorio
5, I-35122 Padova, Italy \and Telescopio Nazionale Galileo, vicolo dell'Osservatorio
5, I-35122 Padova, Italy \and Osservatorio Astronomico di Padova, vicolo dell'Osservatorio
5, I-35122 Padova, Italy \and Instituto de Astrof\`{\i}sica de Canarias, Via
Lactea, E-38200 La Laguna, Tenerife, Spain}

\date{Received / Accepted}

\titlerunning{The red giant branches of Galactic globular clusters}

\maketitle
\begin{abstract}
The purpose of this study is to carry out a thorough investigation of the changes
in morphology of the red giant branch (RGB) of Galactic globular clusters (GGC)
as a function of metallicity, in the \( V,I \) bands. To this aim, two key
points are developed in the course of the analysis.

\textbf{(a)} Using our photometric \( V,I \) database for Galactic globular
clusters (the largest homogeneous data sample to date; Rosenberg et al. \cite{rsp99})
\emph{we measure a complete set of metallicity indices}, based on the morphology
and position of the red-giant branch. In particular, we provide here the first
calibration of the \( S \), \( \Delta V_{1.1} \) and \( \Delta V_{1.4} \)
indices in the \( (V-I,V) \) plane. We show that our indices are internally
consistent, and we calibrate each index in terms of metallicity, both on the
Zinn \& West (1984) and the Carretta \& Gratton (1997) scales. Our new calibrations
of the \( (V-I)_{0,\rm g} \), \( \Delta V_{1.2} \) , \( (V-I)_{-3.0} \) and
\( (V-I)_{-3.5} \) indices are consistent with existing relations.

\textbf{(b)} Using a grid of selected RGB fiducial points, \emph{we define a
function in the \( (V-I)_{0},M_{I},\rm [Fe/H] \) space which is able to reproduce
the whole set of GGC giant branches in terms of a single parameter} (the metallicity).
As a first test, we show that the function is able to predict the correct trend
of our observed indices with metallicity. 

The usage of this function will improve the current determinations of metallicity
and distances within the Local Group, since it allows to easily map \( (V-I)_{0},M_{I} \)
coordinates into \( [{\rm Fe/H}],M_{I} \) ones. To this aim the ``synthetic''
RGB distribution is generated both for the currently used Lee et al. (1990)
distance scale, and for the most recent results on the RR~Lyr distance scale.

\keywords{Stars: abundances - Stars: Population II - Galaxies: abundances -
Globular clusters: general}
\end{abstract}

\section{Introduction \label{s:intro}}

\begin{table}

\caption{The input parameters for the observational sample\label{t:the-sample}}

\begin{tabular}{rrrrrrr}
\hline
\hline\noalign{\smallskip}
                    & & & \multicolumn{3}{c}{$[\rm Fe/H]$} &   \\
\multicolumn{1}{c}{ NGC               }  &
\multicolumn{1}{c}{ $E_{(B-V)}$         }  &
\multicolumn{1}{c}{ $E_{(V-I)}$         }  &
\multicolumn{1}{c}{ ZW                }  &
\multicolumn{1}{c}{ CG                }  &
\multicolumn{1}{c}{ RHS97               }  &
\multicolumn{1}{c}{ $V_{\rm HB}$      }  \\
\noalign{\smallskip}
\hline\noalign{\smallskip}
       104  &     0.05  &     0.06  &    -0.71  &    -0.70  &    -0.78  &    14.05  $\pm$ 0.05 \\
       288  &     0.03  &     0.04  &    -1.40  &    -1.07  &    -1.14  &    15.40  $\pm$ 0.05 \\
       362  &     0.05  &     0.06  &    -1.33  &    -1.15  &    -1.09  &    15.51  $\pm$ 0.05 \\
      1261  &     0.01  &     0.01  &    -1.32  &     ---   &    -1.08  &    16.68  $\pm$ 0.05 \\
      1851  &     0.02  &     0.03  &    -1.23  &     ---   &    -1.03  &    16.18  $\pm$ 0.05 \\
\noalign{\smallskip}								    
      1904  &     0.01  &     0.01  &    -1.67  &    -1.37  &    -1.37  &    16.15  $\pm$ 0.05 \\
      3201  &     0.21  &     0.27  &    -1.53  &    -1.23  &    -1.24  &    14.75  $\pm$ 0.05 \\
      4590  &     0.04  &     0.05  &    -2.11  &    -1.99  &    -2.00  &    15.75  $\pm$ 0.10 \\
      4833  &     0.33  &     0.42  &    -1.92  &    -1.58  &    -1.71  &    15.70  $\pm$ 0.10 \\
      5272  &     0.01  &     0.01  &    -1.66  &     ---   &    -1.33  &    15.58  $\pm$ 0.05 \\
\noalign{\smallskip}								    
      5466  &     0.00  &     0.00  &    -2.22  &     ---   &    -2.13  &    16.60  $\pm$ 0.05 \\
      5897  &     0.08  &     0.10  &    -1.93  &    -1.59  &    -1.73  &    16.30  $\pm$ 0.10 \\
      5904  &     0.03  &     0.04  &    -1.38  &    -1.11  &    -1.12  &    15.00  $\pm$ 0.05 \\
      6093  &     0.18  &     0.23  &    -1.75  &     ---   &    -1.47  &    16.25  $\pm$ 0.05 \\
      6171  &     0.33  &     0.42  &    -1.09  &     ---   &    -0.95  &    15.65  $\pm$ 0.05 \\
\noalign{\smallskip}								    
      6205  &     0.02  &     0.03  &    -1.63  &    -1.39  &    -1.33  &    14.95  $\pm$ 0.10 \\
      6218  &     0.19  &     0.24  &    -1.40  &     ---   &    -1.14  &    14.70  $\pm$ 0.10 \\
      6254  &     0.28  &     0.36  &    -1.55  &    -1.41  &    -1.25  &    15.05  $\pm$ 0.10 \\
      6341  &     0.02  &     0.03  &    -2.24  &     ---   &    -2.10  &    15.20  $\pm$ 0.10 \\
      6352  &     0.21  &     0.27  &    -0.50  &    -0.64  &    -0.70  &    15.25  $\pm$ 0.05 \\
\noalign{\smallskip}								    
      6362  &     0.09  &     0.12  &    -1.18  &    -0.96  &    -0.99  &    15.35  $\pm$ 0.05 \\
      6397  &     0.18  &     0.23  &    -1.94  &    -1.82  &    -1.76  &    12.95  $\pm$ 0.10 \\
      6541  &     0.12  &     0.15  &    -1.79  &     ---   &    -1.53  &    15.40  $\pm$ 0.10 \\
      6637  &     0.17  &     0.22  &    -0.72  &     ---   &    -0.78  &    15.95  $\pm$ 0.05 \\
      6656  &     0.34  &     0.44  &    -1.75  &     ---   &    -1.41  &    14.25  $\pm$ 0.10 \\
\noalign{\smallskip}								    
      6681  &     0.07  &     0.09  &    -1.64  &     ---   &    -1.35  &    15.70  $\pm$ 0.05 \\
      6723  &     0.05  &     0.06  &    -1.12  &     ---   &    -0.96  &    15.45  $\pm$ 0.05 \\
      6752  &     0.04  &     0.05  &    -1.54  &    -1.42  &    -1.24  &    13.80  $\pm$ 0.10 \\
      6779  &     0.20  &     0.26  &    -1.94  &     ---   &    -1.61  &    16.30  $\pm$ 0.05 \\
      6809  &     0.07  &     0.09  &    -1.80  &     ---   &    -1.54  &    14.45  $\pm$ 0.10 \\
\noalign{\smallskip}								    
      7078  &     0.09  &     0.12  &    -2.13  &    -2.12  &    -2.02  &    15.90  $\pm$ 0.05 \\
\noalign{\smallskip}\hline
\end{tabular}

\end{table}
\begin{figure*}
{\par\centering \resizebox*{1\textwidth}{!}{\includegraphics{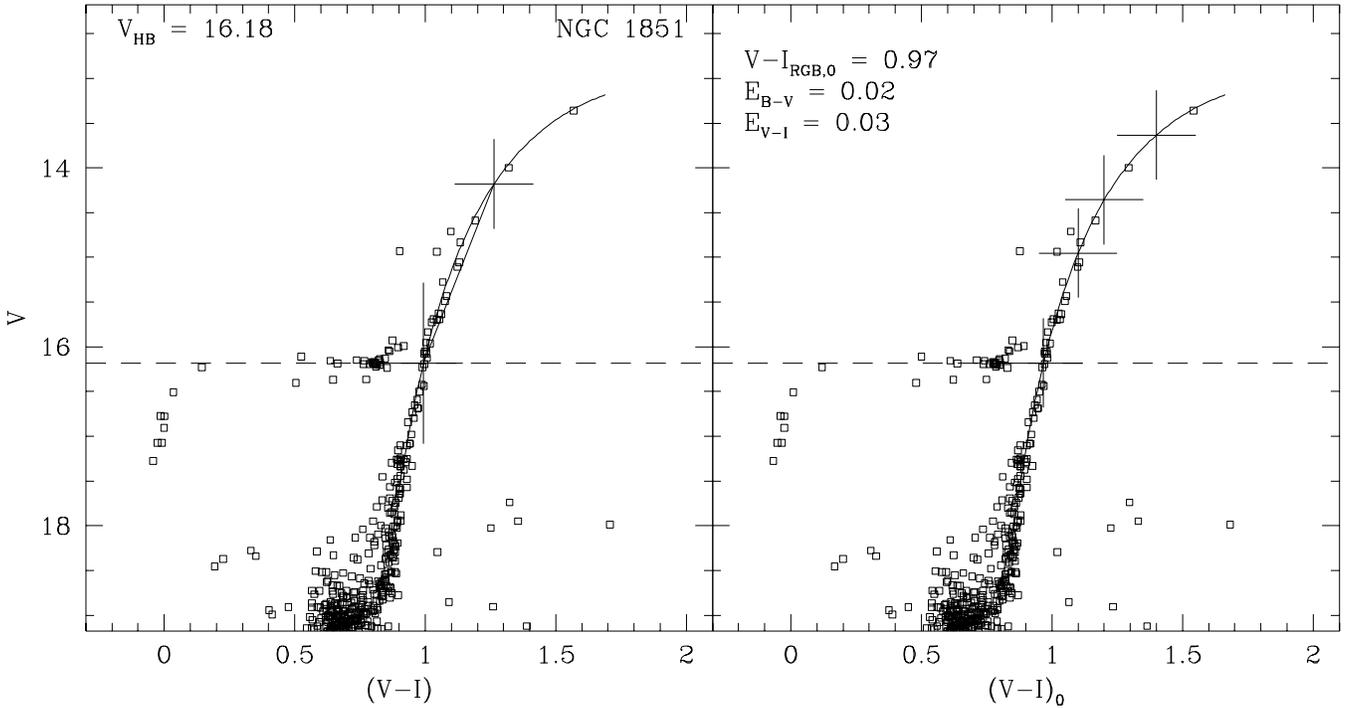}} \par}

\caption{Graphical representation of the metallicity indices (part 1) that were measured
on the selected clusters. (\emph{Left}) The observed CMD of the intermediate-metallicity
cluster NGC1851 and its fiducial RGB (solid line). The fiducial locus was
obtained by fitting Eq.~(\ref{e:iperbole}) to the data.
The two crosses mark the color of the RGB at the level of the HB, and its
color \dvHartwick\ mags brighter than the HB. The slope of the line
connecting the two points is the \protect\( S\protect \)
index. (\emph{Right}) On the color \emph{de-reddened} CMD other four
indices are marked. From fainter to brighter magnitudes, the RGB color at
the level of the HB, and the \protect\( V\protect \) magnitude difference
between this point and those at \protect\( (V-I)_{0}=1.1\protect \), 1.2
and 1.4. \label{f:pars-parteA}.  The dashed line represents the adopted HB
level, \protect\( V_{\rm HB}=16.18 \protect \)}
\end{figure*}
In very recent times, new determinations of Galactic globular cluster (GGC)
metallicities have provided us with new homogeneous \( [\textrm{Fe}/\textrm{H}]\textrm{ } \)
scales. In particular, Carretta \& Gratton (\cite{cg97}; CG) obtained metallicities
from high resolution spectroscopy for 24 GGCs, with an internal uncertainty
of 0.06 dex. For an even larger sample of 71 GGCs, metallicities have been obtained
by Rutledge et al. (\cite{rutledge97}; RHS97) based on spectroscopy of the
Ca\noun{ii} infrared triplet. The equivalent widths of the Ca\noun{ii} triplet have
been calibrated by RHS97 on both the CG scale and the older Zinn \& West (\cite{zw84};
ZW) scale. The compilation by RHS97 is by far the most homogeneous one which
is currently available. 

In the same period, we have been building the largest homogeneous \( V,I \)
photometric sample of Galactic globular clusters (GGC) based on CCD imaging
carried out both with Northern (Isaac Newton Group, ING) and Southern (ESO)
telescopes (Rosenberg et al. \cite{alf3}, \cite{alf4}). The main purpose of
the project is to establish the relative age ranking of the clusters, based
on the methods outlined in Saviane et al. (\cite{srp97}, \cite{srp99}; SRP97,
SRP99) and Buonanno et al. (\cite{b98}; B98). The results of this investigation
are presented in Rosenberg et al. (\cite{rsp99}; RSPA99). Here suffice it to
say that for a set of \reduced\ clusters we obtained \( V \) vs. \( (V-I) \)
color-magnitude diagrams (CMD), which cover a magnitude range that goes from
a few mags below the turnoff (TO) up to the tip of the red giant branch (RGB). 

At this point both a spectroscopic and photometric homogeneous databases are
available: the purpose of this study is to exploit them to perform a thorough
analysis of the morphology of the RGB as a function of the cluster's metallicity.
As a first step, we want to obtain a new improved calibration of a few classical
photometric metallicity indices. Secondly, we want to provide to the community
a self-consistent, \textbf{analytic,} family of giant branches, which can be
used in the analysis of old stellar populations in external galaxies.

\subsection{Metallicity indices}

Photometric indices have been widely used in the past to estimate the mean metallicities
of those stellar systems where direct determinations of their metal content
are not feasible. In particular, they are used to obtain \( [\textrm{Fe}/\textrm{H}]\textrm{ } \)
values for the farthest globulars and for those resolved galaxies of the Local
Group where a significant Pop~II is present (e.g. the dwarf spheroidal galaxies). 

The calibration of \( V,I \) indices is particularly important, since with
comparable exposure times, deeper and more accurate photometry can be obtained
for the cool, low-mass stars in these broad bands than in \( B,V \). Moreover,
our huge CMD database allows a test of the new CG scale on a large basis: we
are able to compare the relations obtained for both the old ZW and new scale,
and check which one allows to rank GGCs in the most accurate way. Indeed, the
most recent calibration of the \( V,I \) indices (Carretta \& Bragaglia \cite{cb98})
is based on just 8 clusters.

\subsection{Old stellar populations in Local Group galaxies}

A reliable metallicity ranking of GGC giant branches also allows studies
that go beyond a simple determination of the \emph{mean} metallicity of a
stellar population. As an illustration, we may recall the recent
investigation of the halo metallicity distribution function (MDF) of
NGC~5128 (Harris et al.~\cite{harris5128}), which was based on the fiducial
GC lines obtained by Da Costa \& Armandroff (\cite{da90}, hereafter
DA90). These studies can be made more straightforward by providing a
suitable analytic representation of the RGB family of GGCs. Indeed,
assuming that most of the GGCs share a common age (e.g. Rosenberg et
al. \cite{rsp99}), one expects that there should exist a ``universal''
function of \( \{(V-I)_{0},M_{I},\rm [Fe/H]\} \) able to map any \(
[(V-I)_{0},M_{I}] \) coordinate pair into the corresponding metallicity
(provided that an independent estimate of the distance and extinction of
the star are available). We will show here that such relatively simple
mono-parametric function can actually be obtained, and that this progress is
made possible thanks to the homogeneity of both our data set and analysis.

In order to enforce a proper use of our calibrations, we must clearly state
that, in principle, the present relations are valid only for rigorously old
stellar populations (i.e. for stars as old as the bulk of Galactic
globulars). At fixed abundance, giant branches are somewhat bluer for
younger ages (e.g. Bertelli et al. \cite{bertelli94}).  Moreover, in real
stellar systems AGB stars are also present on the blue side of the RGB
(cf. Fig.~\ref{f:pars-partB}). Both effects must be taken into account when
dealing with LG galaxies, since they could lead to systematic effects in
both the mean abundances and the abundance distributions (e.g. Saviane et
al. \cite{fornax}).

\subsection{Layout of the paper}

The observational sample, on which this investigation is based, is presented
in Sect.~\ref{s:sample}. Sect.~\ref{s:indices} is devoted to the set of indices
which are to be calibrated. They are defined in Sect~\ref{s:defindices}. The
reliability of our sample is tested in Sect.~\ref{s:checks}, where we demonstrate
that our methodology produces a set of well-correlated indices. In Sect.~\ref{s:newda90}
we show that, once a distance scale is assumed for the GGCs, our whole set of
RGBs can be approximated by a \emph{single} analytic function, which depends
on the metallicity alone. This finding allows a new and easier way to determine
the distances and mean metallicities of the galaxies of the Local Group, extending
the methods of Da Costa \& Armandroff (\cite{da90}), and Lee et al. (\cite{lee.et.al93}).
The metallicity indices are calibrated in Sect.~\ref{s:calibrations}, where
analytic relations are provided both for the ZW and for the CG scales. Using
these indices, we are able to test our analytic RGB family in Sect~\ref{s:testfits}.
Our conclusions are in Sect.~\ref{s:conclusioni}.

\section{The observational sample \label{s:sample}}

\nDutchText\ clusters have been observed with the ESO/Dutch 0.9m telescope at
La Silla, and \nJKT\ at the RGO/JKT 1m telescope in la Palma. This database
comprises \( 75\% \) of the GGC whose distance modulus is \( (m-M)_{\rm V}<\mMUnMetro  \).
The zero-point uncertainties of our calibrations are \( <0.03 \)~mag for each
band. Three clusters were observed both with the southern and the northern telescopes,
thus providing a consistency check of the calibrations: no systematic differences
were found, at the level of accuracy of the zero-points. A detailed description
of the observations and reduction procedures will be given in forthcoming papers
(Rosenberg et al. \cite{alf3}, \cite{alf4}) presenting the single clusters. 

\begin{figure}
{\par\centering \resizebox*{1\columnwidth}{!}{\includegraphics{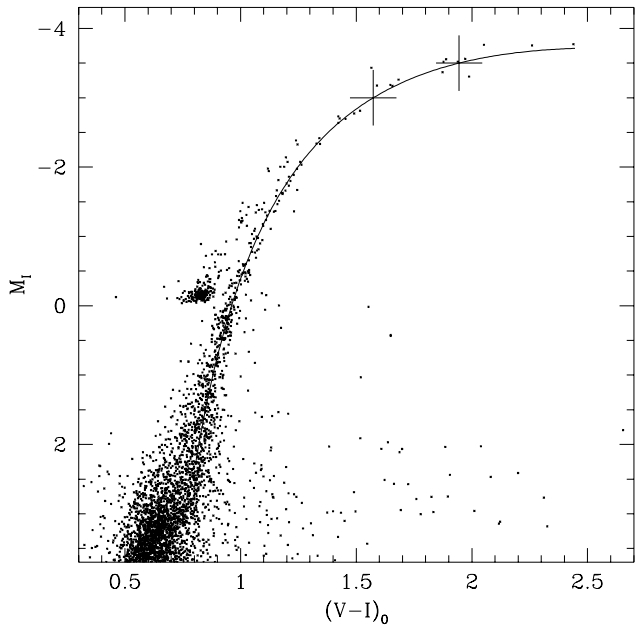}} \par}

\caption{Graphical representation of the metallicity indices (part 2) that were measured
on the selected clusters. In this figure, the \emph{absolute} CMD of the metal
rich cluster NGC 104 is plotted in the \protect\( (V-I)_{0},M_{I}\protect \)
plane, adopting an apparent distance modulus \protect\( (m-M)_{V}=13.35\protect \)
and a reddening \protect\( E_{V-I}=0.06\protect \) (see text for the discussion).
This plot shows the ability of the analytic function to reproduce even the more
extended RGBs. The two crosses mark the color of the RGB at \protect\( M_{I}=-3.0\protect \)
and \protect\( M_{I}=-3.5\protect \).\label{f:pars-partB}}
\end{figure}

A subsample of this database was used for the present investigation. We retained
those clusters whose CMD satisfied a few criteria: (a) the HB level could be
well determined; (b) the RGB was not heavily contaminated by foreground/background
contamination; and (c) the RGB was well defined up to the tip. This subsample
largely overlaps that used for the age investigation, but a few clusters whose
TO position could not be measured, are nevertheless useful for the metallicity
indices definition. Conversely, in a few cases the lower RGB could be used for
the color measurements, while the upper branch was too scarcely defined for
a reliable definition of the fiducial line. Two of the CMDs that were used are
shown in Figs.~\ref{f:pars-parteA} (NGC~1851) and \ref{f:pars-partB} (NGC~104),
and they illustrate the good quality of the data.

The dataset of \nGCutili\ clusters used in this paper is listed in Table~\ref{t:the-sample}.
From left to right, the columns contain the NGC number, the reddening both in
\( (B-V) \) and \( (V-I) \), the metallicity according to three different
scales, and the apparent magnitude of the horizontal branch (HB). The \( E_{(B-V)} \)
values were taken from the Harris (\cite{harris96}) on-line table\footnote{
http://physun.physics.mcmaster.ca/Globular.html
}. The \( (V-I) \) reddenings were obtained by assuming that \( E_{(V-I)}=1.28\times E_{(B-V)} \)
(Dean et al. \cite{dwc78}). The values of the metallicity were taken from RHS97:
they represent the equivalent widths of the Ca\noun{ii} infrared triplet,
calibrated either onto the Zinn \& West (\cite{zw84}) scale (ZW column) or
the Carretta \& Gratton (\cite{cg97}) scale (RHS97 column). Moreover, the original
Carretta \& Gratton metallicities (CG column) are also given for the clusters
comprised in their sample.

The HB level was found in different ways for clusters of different metallicity.
For the the metal rich and metal intermediate clusters, a magnitude distribution
of the HB stars was obtained, and the mode of the distribution was taken. Where
the HB was too scarcely populated, a horizontal line was fitted through the
data. The blue tail of the metal poorest clusters does not reach the horizontal
part of the branch: in that case, a fiducial HB was fitted to the tail, and
the magnitude of the horizontal part was taken as the reference level. The fiducial
branch was defined by taking a cluster having a bimodal HB color distribution
(NGC~1851, cf. Fig.~\ref{f:pars-parteA}) and then extending its HB both to
the red and to the blue by ``appending'' clusters being more and more metal
rich and metal poor, respectively. The details of this procedure, as well as
the errors associated to the \( V_{\rm HB} \) in Table~\ref{t:the-sample},
are discussed in RSPA99. For NGC~1851, \( V_{\rm HB}=16.18\pm 0.05 \) was adopted
(dashed line in Fig.~\ref{f:pars-parteA}), and this value is just \( 0.02 \)~mag
brighter than the value found by Walker (\cite{walker1851}) and Saviane et
al. (\cite{ivo1851}). 

Based on this observational sample, a set of metallicity indices were measured
on the RGBs of the clusters. In the next section, the indices are defined and
the measurement procedures are described. Consistency checks are also performed.

\section{Metallicity indices \label{s:indices}}

\subsection{Definitions\label{s:defindices}}

The metallicity indices calibrated in this study are represented and defined
in Fig.~\ref{f:pars-parteA} and Fig.~\ref{f:pars-partB}.  The figures represent
the CMD of NGC~1851 and NGC~104 in different color-magnitude planes, and the
crosses mark the position of the RGB points used in the measurement of the indices. 

The left panel of Fig.~\ref{f:pars-parteA} shows the apparent colors and magnitudes
for NGC~1851: the inclined line helps to identify the first index, \( S \).
This was defined, in the \( (B-V,V) \) plane, by Hartwick (\cite{hartwick68})
as the slope of the line connecting two points on the RGB: the first one at
the level of the HB, and the second one 2.5 mag brighter. We use the same definition
for the \( (V-I,V) \) plane here; however, in order to be able to use our metal
richest clusters, we redefined \( S \) by measuring the second RGB point \dvHartwick\
mag brighter than the HB. Since \( S \) is measured on the apparent CMD, it
is independent both from the reddening and the distance modulus. 

The right panel of the same figure, shows the apparent \( V \) magnitude vs.
the de-reddened \( (V-I)_{0} \) color. In this panel, four other indices are
identified, i.e. \( (V-I)_{0,g} \), \( \Delta V_{1.1} \), \( \Delta V_{1.2} \),
and \( \Delta V_{1.4} \). The first one is the RGB color at the level of the
HB, and the other three measure the magnitude difference between the HB and
the RGB at a fixed color \( (V-I)_{0}=1.1 \), 1.2 and 1.4 mag. The former index
was originally defined by Sandage \& Smith (\cite{sandsmith66}) and the latter
one by Sandage \& Wallerstein (\cite{sandwall60}), in the \( (B-V)_{0},V \)
plane. The other two indices, \( \Delta V_{1.1} \) and \( \Delta V_{1.2} \),
are introduced later to measure the metal richest GCs. These indices require
an independent color excess determination. 

Finally, Fig.~\ref{f:pars-partB} shows the CMD of NGC~104 (47~Tuc) in the absolute
\( (V-I)_{0,}M_{I} \) plane: the adopted distance modulus, \( (m-M)_{V}=13.35 \),
was obtained by correcting the apparent luminosity of the HB according to
Lee et al. (\cite{ldz}; cf. Sect.~\ref{s:calibrations}). By comparison,
Harris' catalog reports \( (m-M)_{V}=13.32 \). Two other indices are
represented in the figure: \( (V-I)_{-3.0} \) and \( (V-I)_{-3.5} \).  They
are defined as the RGB color at a fixed absolute \( I \) magnitude of \(
M_{I}=-3.0 \) (Da Costa \& Armandroff \cite{da90}) or \( M_{I}=-3.5 \) (Lee
et al. \cite{lee.et.al93}).
The latter index was also discussed by Armandroff et
al. (\cite{taft-3p5}), and a calibration formula was given in Caldwell
et al. (\cite{caldwell-3p5}). This is based on the DA90 clusters plus M5
and NGC~362 from Lloyd Evans (\cite{lloyd83}).

Since these two indices are defined on the bright part of the RGB,
they can be measured even for the farthest objects of the Local Group
(LG). Due to the fast luminosity evolution of the stars on the upper RGB,
this part of the branch was typically under-sampled by the early small-size
CCDs, so no wide application of these indices has been made for Galactic
globulars. However, this is of no concern for galaxy-size stellar
systems. It will be shown in Sect.~\ref{s:calibrations} that good
accuracies can be obtained even for GCs, provided that the analytic
function of Eq.~(\ref{e:iperbole}) is used.

\subsection{Measurement procedures}

\begin{table}

\caption{The measured metallicity indices \label{t:the-indices}}

\begin{tabular}{lrrrrrrr}
\hline
\hline\noalign{\smallskip}
        &      &       &  \multicolumn{3}{c}{$\Delta V$}  & \multicolumn{2}{c}{$(V-I)$}     \\
\multicolumn{1}{c}{NGC}                 &
\multicolumn{1}{c}{$(V-I)_{0,\rm g}$}     &
\multicolumn{1}{c}{$S$}                 &
\multicolumn{1}{c}{$1.1$}    &
\multicolumn{1}{c}{$1.2$}    &
\multicolumn{1}{c}{$1.4$}    &
\multicolumn{1}{c}{$@3.5$}         &
\multicolumn{1}{c}{$@3.0$} \\
\noalign{\smallskip}
\hline\noalign{\smallskip}
       104  &     0.99  &     4.13  &     0.78  &     1.27  &     1.87  &     1.94  &     1.57 \\
       288  &     0.95  &     6.39  &     1.25  &     1.75  &     2.36  &     1.51  &     1.35 \\
       362  &     0.90  &     7.28  &     1.67  &     2.09  &     2.57  &     1.45  &     1.28 \\
      1261  &     0.91  &     7.77  &     1.62  &     2.13  &     2.73  &     1.39  &     1.25 \\
      1851  &     0.97  &     7.41  &     1.23  &     1.82  &     2.55  &     1.45  &     1.31 \\
\noalign{\smallskip}
      1904  &     0.94  &     8.56  &     1.58  &     2.14  &     2.83  &     1.35  &     1.24 \\
      3201  &     0.99  &     8.72  &     1.19  &     1.91  &     2.71  &     1.39  &     1.27 \\
      4590  &     0.91  &     9.98  &     1.90  &     2.52  &     3.25  &     1.24  &     1.16 \\
      4833  &     0.92  &     9.25  &     1.80  &     2.36  &     3.12  &     1.28  &     1.19 \\
      5272  &     0.91  &     7.60  &     1.66  &     2.13  &     2.81  &     1.36  &     1.24 \\
\noalign{\smallskip}
      5466  &     0.91  &     9.85  &     1.93  &     2.50  &     3.18  &     1.24  &     1.16 \\
      5897  &     0.97  &     8.73  &     1.34  &     2.00  &     2.79  &     1.35  &     1.25 \\
      5904  &     0.93  &     6.91  &     1.41  &     1.91  &     2.55  &     1.44  &     1.30 \\
      6093  &     0.93  &     8.02  &     1.58  &     2.12  &     2.91  &     1.34  &     1.24 \\
      6171  &     1.07  &     5.66  &     0.31  &     1.09  &     1.93  &     1.67  &     1.49 \\
\noalign{\smallskip}
      6205  &     0.89  &     7.70  &     1.75  &     2.20  &     2.75  &     1.37  &     1.23 \\
      6218  &     0.95  &     7.09  &     1.34  &     1.88  &     2.51  &     1.46  &     1.31 \\
      6254  &     0.90  &     8.25  &     1.75  &     2.29  &     3.17  &     1.30  &     1.21 \\
      6341  &     0.88  &     9.92  &     2.15  &     2.69  &     3.40  &     1.21  &     1.13 \\
      6352  &     1.12  &     3.11  &    -0.16  &     0.52  &     1.30  &     1.99  &     1.75 \\
\noalign{\smallskip}
      6362  &     0.93  &     5.84  &     1.31  &     1.76  &     2.32  &     1.55  &     1.37 \\
      6397  &     0.89  &     9.45  &     1.98  &     2.49  &     3.12  &     1.26  &     1.16 \\
      6541  &     1.01  &     8.59  &     1.03  &     1.77  &     2.67  &     1.39  &     1.29 \\
      6637  &     0.96  &     4.39  &     0.96  &     1.41  &     1.97  &     1.82  &     1.53 \\
      6656  &     0.86  &    10.32  &     2.27  &     2.69  &     2.96  &     1.24  &     1.12 \\
\noalign{\smallskip}
      6681  &     0.95  &     7.54  &     1.35  &     1.92  &     2.76  &     1.37  &     1.27 \\
      6723  &     1.01  &     6.02  &     0.76  &     1.38  &     2.18  &     1.55  &     1.41 \\
      6752  &     0.99  &     7.16  &     1.08  &     1.69  &     2.46  &     1.45  &     1.33 \\
      6779  &     0.94  &     8.74  &     1.60  &     2.18  &     2.94  &     1.32  &     1.22 \\
      6809  &     0.93  &     9.38  &     1.72  &     2.29  &     2.87  &     1.32  &     1.20 \\
\noalign{\smallskip}
      7078  &     0.88  &     9.82  &     2.10  &     2.62  &     3.27  &     1.23  &     1.14 \\
\noalign{\smallskip}
\hline
\end{tabular}

\end{table}

Colors and magnitudes were measured on a fiducial RGB, which has been found
by least-square fitting an analytic function to the observed branch.  After
some experimenting, it was found that the best solution is to use the following
relation:   
\begin{equation}
\label{e:iperbole}
y=a+bx+c/(x-d)
\end{equation}
 where \( x \) and \( y \) represent the color and the magnitude, respectively.
One can see from Figs.~\ref{f:pars-parteA} and \ref{f:pars-partB} that the
function is indeed able to represent the giant branch over the typical metallicity
range of globular clusters. Moreover, it is shown in Sect.~\ref{s:newda90}
that, when the CMDs are corrected for distance and reddening, the four coefficients
can be parametrized as a function of {[}Fe/H{]}, so that one is able to reproduce
the RGB of each cluster, using just one parameter: the metallicity. At any rate,
the indices were measured on the original loci, so that an independent check
of the goodness of the generalized hyperbolae can be made, by comparison of
the measured vs. predicted indices.

All the indices' values that have been measured are reported in Table~\ref{t:the-indices}.
In this table, the cluster NGC number is given in column 1; the following columns
list, from left to right, \( (V-I)_{0,\rm g} \), \( S \), \( \Delta V_{1.1} \),
\( \Delta V_{1.2} \), \( \Delta V_{1.4} \), and finally the RGB color measured
at \( M_{I}=-3 \) and \( -3.5 \). The Lee et al. (1990) distance scale was
used to compute the last two indices (cf. Sect.~\ref{s:calibrations}).

\subsection{Internal consistency checks \label{s:checks}}

\begin{figure}
{\par\centering \resizebox*{1\columnwidth}{!}{\includegraphics{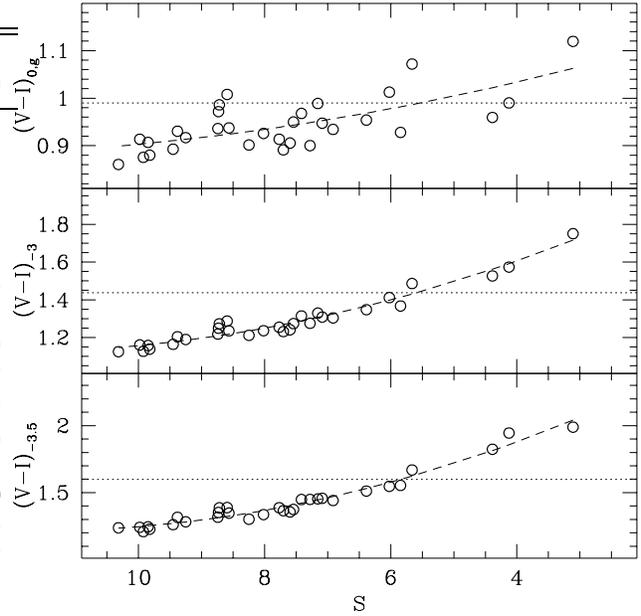}} \par}

\caption{The metallicity indices \protect\( (V-I)_{0,\rm g}\protect \), \protect\( (V-I)_{-3.0}\protect \),
and \protect\( (V-I)_{-3.5}\protect \) are plotted as a function of the index
\protect\( S\protect \). The points are ordered such that the metal-poor to
metal-rich cluster sequence goes from left to right. The very good correlations
between \protect\( (V-I)_{-3.0}\protect \), \protect\( (V-I)_{-3.5}\protect \)
and \protect\( S\protect \) (the \emph{rms} of the parabolic fits are 2\% and
3\% respectively), demonstrate that these indices are very accurate \label{f:intcheck2}}
\end{figure}
\begin{figure}
{\par\centering \resizebox*{1\columnwidth}{!}{\includegraphics{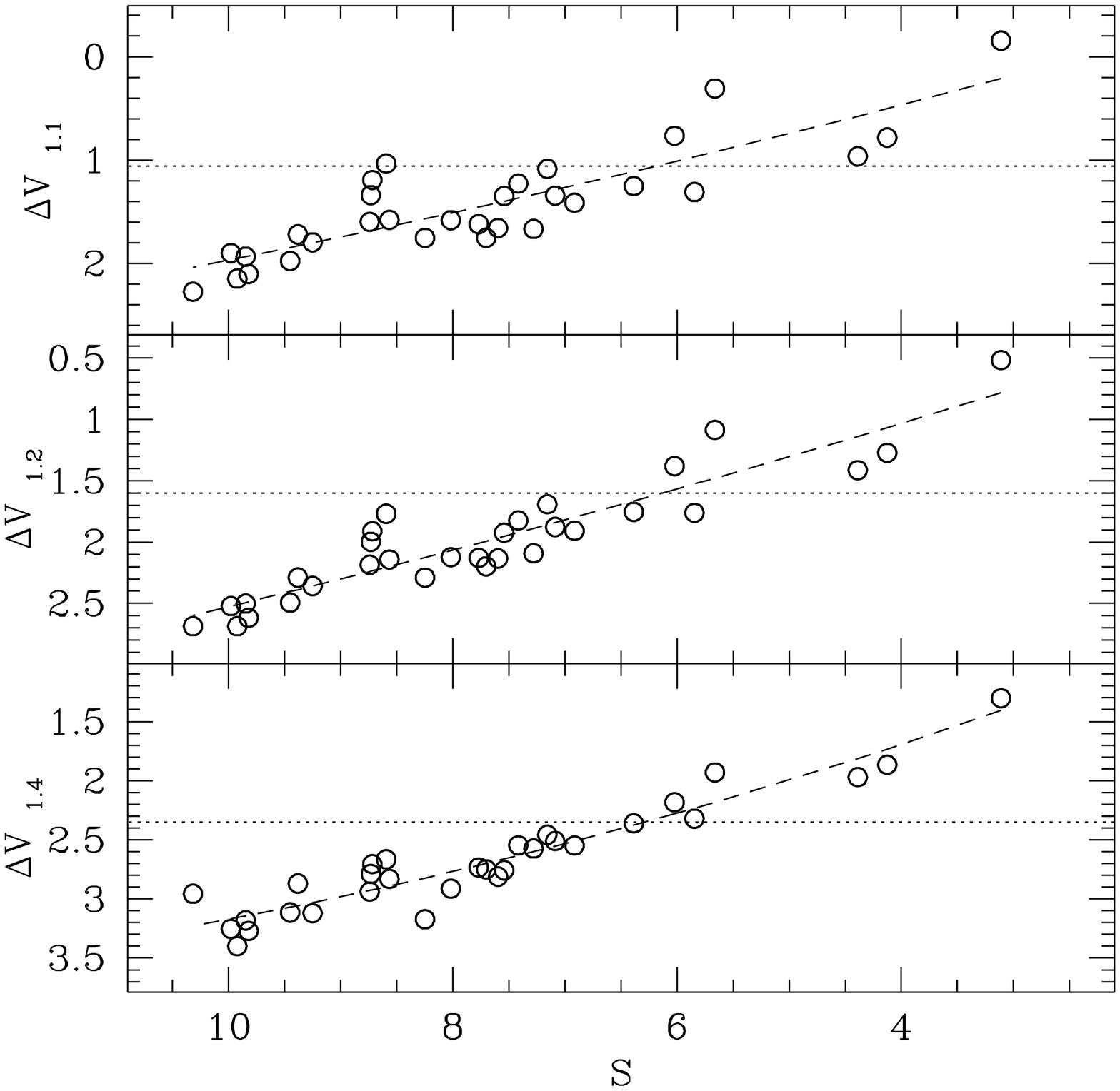}} \par}

\caption{The metallicity indices \protect\( \Delta V_{1.1}\protect \) , \protect\( \Delta V_{1.2}\protect \)
, and \protect\( \Delta V_{1.4}\protect \) as a function of the index \protect\( S\protect \).
The points are ordered such that the metal-poor to metal-rich cluster sequence
goes from left to right. The dashed line represents a second order polynomial
least-square fitted to the data. The typical relative uncertainty on each index
has been estimated by taking the ratio of the \emph{rms} of the fit over the
value of the parameter at the level of the horizontal dotted line\label{f:intcheck1}}
\end{figure}

Before discussing the indices as metallicity indicators, we checked their internal
consistency. We will show in Sect.~\ref{s:calibrations} that the index \( S \)
is the most accurate one, as expected, since it does not require reddening and
distance corrections. The rest of the indices are therefore plotted vs. \( S \)
in Figs.~\ref{f:intcheck2} and \ref{f:intcheck1}, and we expect that most
of the scatter will be in the vertical direction. Second order polynomials were
fitted to the distributions, and the \emph{rms} of the fit was computed for
each index. In order to intercompare the different indices, a relative uncertainty
has been computed by dividing the \emph{rms} by the central value of each parameter
(this value is identified by a dotted line in each figure). 

In this way, the scatter of the metal index \( i \) is \( \Delta i/i=0.02 \),
0.02, 0.04, 0.06, 0.12, and 0.26, for the indices \( (V-I)_{-3} \), \( (V-I)_{-3.5} \),
\( (V-I)_{0,\rm g} \), \( \Delta V_{1.4} \), \( \Delta V_{1.2} \), and \( \Delta V_{1.1} \),
respectively.
These values confirm the visual impression of the figures, that \( \Delta (V-I)_{-3.0} \)
and \( \Delta (V-I)_{-3.5} \) are the lowest dispersion indices, followed by
\( (V-I)_{0,\rm g} \) and \( \Delta V_{1.4} \).

The indices will be calibrated in terms of {[}Fe/H{]} in Sect.~\ref{s:calibrations};
however, before moving to this section, we want to present a new way to provide
``standard'' GGC branches in the \( (V-I)_{0},M_{I} \) plane, along the lines
of the classical Da Costa \& Armandroff (\cite{da90}) study. Using this family
of RGB branches, we are able to make predictions on the trend of the already
defined indices with metallicity; these trends can thus be compared to the observed
ones, and therefore provide a further test of the reliability of our RGB family
(cf. Sect~\ref{s:testfits}).

\section{New standard globular cluster giant branches \label{s:newda90}}

\begin{table*}

\caption{Clusters selected for the determination of the analytic fits,
listed for increasing [Fe/H] values \label{t:fiducialGC}}

\begin{tabular}{lcccccc}
\hline\hline\noalign{\smallskip}
Cluster & $V_{\rm HB}$ & $E_{(B-V)}$ & $E_{(V-I)}$ & [Fe/H]$_{\rm ZW}$ & [Fe/H]$_{\rm RCG}$ &[Fe/H]$_{\rm CG}$ \\
\noalign{\smallskip}\hline
 NGC 104  & 14.05 & 0.050 & 0.064 & $-0.71$ & $-0.78$ & $-0.70$  \\ 
 NGC 5904 & 15.00 & 0.023 & 0.029 & $-1.38$ & $-1.12$ & $-1.11$  \\ 
 NGC 288  & 15.40 & 0.036 & 0.046 & $-1.40$ & $-1.14$ & $-1.07$  \\ 
 NGC 6205 & 14.95 & 0.000 & 0.000 & $-1.63$ & $-1.33$ & $-1.39$  \\ 
 NGC 5272 & 15.58 & 0.002 & 0.003 & $-1.66$ & $-1.33$ &   ---  \\ 
 NGC 6341 & 15.20 & 0.010 & 0.013 & $-2.24$ & $-2.10$ &   ---  \\ 
\noalign{\smallskip}\hline
\end{tabular}

\end{table*}
\begin{table*}

\caption{The fiducial points for the 6 selected clusters \label{t:fidtable}}

\begin{tabular}{rrrrrrrrrrrr} 
\hline \hline
\noalign{\smallskip}
\multicolumn{2}{c}{NGC 104 }&
\multicolumn{2}{c}{NGC 288 }&
\multicolumn{2}{c}{NGC 5272 }&
\multicolumn{2}{c}{NGC 5904 }&
\multicolumn{2}{c}{NGC 6205 }&
\multicolumn{2}{c}{NGC 6341 }\\
\noalign{\smallskip}
   $I$  & $(V-I)$ &   $I$  & $(V-I)$ &   $I$  & $(V-I)$ &   $I$  & $(V-I)$ &   $I$  & $(V-I)$ &   $I$  & $(V-I)$ \\
\noalign{\smallskip}
\hline
\noalign{\smallskip}
 13.782 & 0.978 & 15.359 & 0.914 & 15.492 & 0.852 & 14.725 & 0.926 & 14.645 & 0.867 & 15.060 & 0.852  \\ 
 13.604 & 0.994 & 15.107 & 0.939 & 15.151 & 0.874 & 14.457 & 0.942 & 14.322 & 0.890 & 14.720 & 0.872  \\ 
 13.443 & 1.008 & 14.849 & 0.960 & 14.789 & 0.892 & 14.221 & 0.961 & 14.033 & 0.909 & 14.395 & 0.894  \\ 
 13.317 & 1.021 & 14.593 & 0.984 & 14.597 & 0.910 & 14.040 & 0.978 & 13.788 & 0.929 & 14.079 & 0.916  \\ 
 13.075 & 1.045 & 14.342 & 0.999 & 14.359 & 0.929 & 13.878 & 0.994 & 13.595 & 0.944 & 13.789 & 0.937  \\ 
 12.862 & 1.070 & 14.109 & 1.018 & 14.143 & 0.955 & 13.700 & 1.009 & 13.381 & 0.966 & 13.533 & 0.953  \\ 
 12.619 & 1.101 & 13.881 & 1.036 & 13.796 & 0.990 & 13.456 & 1.032 & 13.170 & 0.984 & 13.303 & 0.974  \\ 
 12.346 & 1.136 & 13.649 & 1.062 & 13.517 & 1.021 & 13.190 & 1.061 & 12.984 & 1.005 & 13.082 & 0.994  \\ 
 12.035 & 1.185 & 13.376 & 1.090 & 13.265 & 1.046 & 12.916 & 1.091 & 12.832 & 1.019 & 12.850 & 1.020  \\ 
 11.761 & 1.231 & 13.058 & 1.132 & 13.005 & 1.076 & 12.655 & 1.122 & 12.631 & 1.045 & 12.611 & 1.039  \\ 
 11.461 & 1.281 & 12.766 & 1.173 & 12.759 & 1.110 & 12.419 & 1.154 & 12.363 & 1.077 & 12.351 & 1.067  \\ 
 11.101 & 1.362 & 12.534 & 1.210 & 12.519 & 1.148 & 12.231 & 1.183 & 12.118 & 1.111 & 12.075 & 1.102  \\ 
 10.696 & 1.459 & 12.380 & 1.233 & 12.302 & 1.187 & 12.073 & 1.212 & 11.945 & 1.138 & 11.771 & 1.148  \\ 
 10.330 & 1.600 & 12.163 & 1.268 & 12.109 & 1.227 & 11.868 & 1.254 & 11.844 & 1.156 & 11.492 & 1.195  \\ 
 10.062 & 1.720 & 11.928 & 1.317 & 11.878 & 1.275 & 11.615 & 1.305 & 11.707 & 1.178 & 11.284 & 1.233  \\ 
 9.877 & 1.856 & 11.617 & 1.411 & 11.741 & 1.310 & 11.335 & 1.371 & 11.571 & 1.204 & 11.154 & 1.265  \\ 
 9.706 & 2.019 & 11.427 & 1.483 & 11.575 & 1.344 & 11.116 & 1.422 & 11.395 & 1.252 & 11.008 & 1.295  \\ 
 9.602 & 2.148 &  ---  &  ---  & 11.494 & 1.377 & 10.902 & 1.489 & 11.141 & 1.312 & 10.854 & 1.320  \\ 
 9.524 & 2.315 &  ---  &  ---  & 11.330 & 1.406 & 10.652 & 1.585 & 10.870 & 1.376 & 10.709 & 1.351  \\ 
 9.573 & 2.576 &  ---  &  ---  & 11.240 & 1.447 & 10.457 & 1.680 & 10.643 & 1.444 & --- & ---  \\ 
 9.619 & 2.768 &  ---  &  ---  & 11.112 & 1.488 & 10.343 & 1.742 & 10.552 & 1.492 & --- & ---  \\ 
  ---  &  ---  &  ---  &  ---  & 11.078 & 1.528 &  ---  &  ---  &  ---  &  ---  &  ---  &  ---   \\ 
  ---  &  ---  &  ---  &  ---  & 11.047 & 1.546 &  ---  &  ---  &  ---  &  ---  &  ---  &  ---   \\ 
\noalign{\smallskip}
\hline
\end{tabular}

\end{table*}

Da Costa \& Armandroff (\cite{da90}) presented in tabular form the fiducial
GGC branches of 6 globulars, covering the metallicity range \( -2.17\leq [\rm Fe/H]\leq -0.71 \).
The RGBs were corrected to the absolute \( (V-I)_{0},M_{I} \) plane using the
apparent \( V \) magnitude of the HB, and adopting the Lee et al. (\cite{ldz})
theoretical HB luminosity. Since the DA90 study, these branches have been widely
used for stellar population studies in the Local Group. Based on these RGBs,
in particular, a method to determine both the distance and mean metallicity
of an old stellar population \noun{}was presented by Lee et al. (\cite{lee.et.al93}). 

Both DA90 and Lee et al. (1993) provided a relation between the metallicity
{[}Fe/H{]} and the color of the RGB at a fixed absolute \( I \) magnitude
(\( M_{I}=-3 \) and \( -3.5 \), respectively),
and recently a new relation for $(V-I)_{-3.5}$ has also been obtained by
Caldwell et al. (\cite{caldwell-3p5}).
Once the distance of the population is known (e.g. via the luminosity of
the RGB tip), then an estimate of its \emph{mean} metallicity can be
obtained using one of the  calibrations. It is assumed that the age of
the population is comparable to that of the GGCs, and that the age spread
is negligible compared to the metallicity spread (RSPA99).

In such case, one expects that any RGB star's position in the absolute CMD is
determined just by its metallicity, and that a better statistical determination
of the population's metal content would be obtained by converting the color
of \emph{each} star into a {[}Fe/H{]} value. With this idea in mind, in the
following sections we will show that this is indeed possible, at least for the
bright/most sensitive part of the giant branch. We found that a relatively simple
\emph{continuous} function can be defined in the \( (V-I)_{0},M_{I},\rm [Fe/H] \)
space, and that this function can be used to transform the RGB from the \( (V-I)_{0},M_{I} \)
plane to the \( {\rm [Fe/H]},M_{I} \) plane. 

In order to obtain this function, we first selected a subsample of clusters
with suitable characteristics, so that a reference RGB grid can be constructed.
The fiducial branches for each cluster were then determined in an objective
way, and they were corrected to the absolute \( ((V-I)_{0},M_{I}) \) plane.
In this plane, the analytic function was fitted to the RGB grid. These operations
are described in the following sections.

\subsection{Selection of clusters}

The clusters that were used for the definition of the fiducial RGBs are
listed in Table~\ref{t:fiducialGC}, in order of increasing metallicity. The
table reports the cluster name, and some of the parameters listed in
Table~\ref{t:the-sample} are repeated here for ease of use. The values of
the reddening were in some cases changed by a few thousandth magnitudes
(i.e. well within the typical uncertainties on \( E_{(B-V)} \)), to obtain a
sequence of fiducial lines that move from bluer to redder colors as
{[}Fe/H{]} increases, and again the corresponding \( E_{(V-I)} \) values
were obtained assuming that \( E_{(V-I)}=1.28\times E_{(B-V)} \) (Dean et
al. \cite{dwc78}). Indeed, due to the homogeneity of our sample, we expect
that if a monotonic color/metallicity sequence is not obtained, then only
the uncertainties on the extinction values must be taken into account.

In order to single out these clusters from the total sample, some key characteristics
were taken into account. In particular, we considered clusters whose RGBs are
all well-defined by a statistically significant number of stars; they have low
reddening values (\( E_{(B-V)}\leq 0.05 \)); and they cover a metallicity range
that includes most of our GGCs (\( -2.2\leq \rm [Fe/H]\leq -0.7 \) on the ZW
scale). 

The DA90 fiducial clusters were NGC~104, NGC~1851, NGC~6752, NGC~6397, NGC~7078
and NGC~7089 (M2). NGC 104 is the only cluster in common with the previous study,
and M2 is not present in our dataset. The other objects have been excluded from
our fiducial sample since they have too large reddening values (\( E_{(B-V)}>0.05 \)
for NGC~6397 and NGC~7078), or their RGBs are too scarcely populated in our
CMDs (NGC 1851 and NGC~6752). Nevertheless, the calibrations that we obtain
for the \( (V-I)_{-3.0} \) and \( (V-I)_{-3.5} \) are in fairly good agreement
with those obtained by DA90 (for the small discrepancies at the high metallicity
end, cf. Sect.~\ref{s:DA90indices1} and \ref{s:DA90indices2}), and in
particular with the recent Caldwell et al. (\cite{caldwell-3p5})
calibration for the $(V-I)_{-3.5}$ index.

\subsection{Determination of the fiducial loci\label{s:fiducials}}

\begin{figure*}
{\par\centering \resizebox*{1.7\columnwidth}{!}{\includegraphics{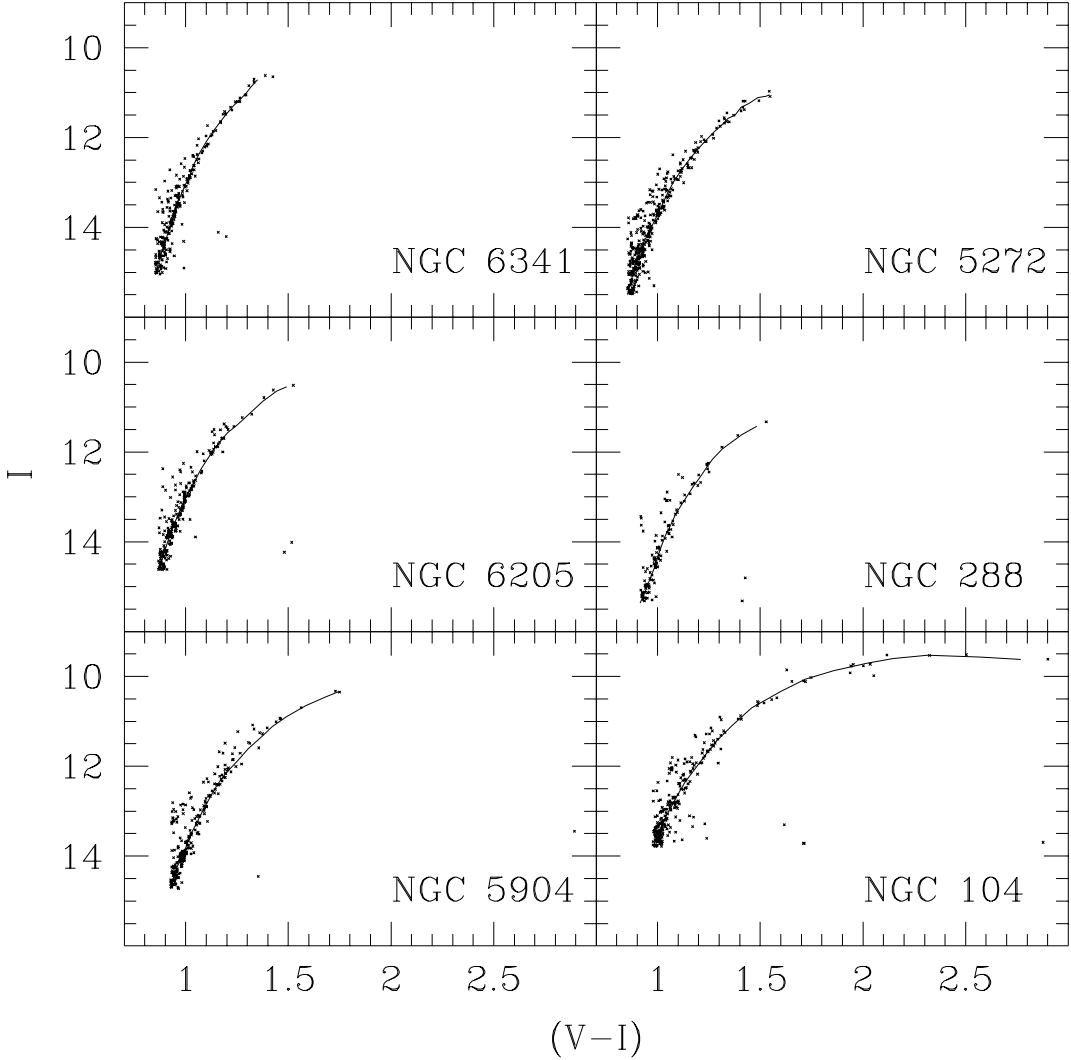}} \par}

\caption{The RGBs of the selected clusters (crosses) and their fiducial lines (solid
curves). The clusters are represented, from left to right and top to bottom,
in order of increasing metallicity \label{f:zooms}}
\end{figure*}

The ridge lines of our fiducial RGBs were defined according to the following
procedure. The RGB region was selected from the calibrated photometry, by excluding
both HB and AGB stars. All stars bluer than the color of the RR~Lyr gap were
removed; AGB stars were also removed by tracing a reference straight line in
the CMD, and by excluding all stars blue-side of this line. This operation was
carried out in the \( ((V-I),I) \) plane, where the RGB curvature is less pronounced,
and a straight line turns out to be adequate. 

The fiducial loci were then extracted from the selected RGB samples. The \( (V-I) \)
and \( I \) vectors were sorted in magnitude, and bins were created containing
a given number of stars. Within each bin, the median color of the stars and
the mean magnitude were used as estimators of the bin central color and brightness.
The number of stars within the bins was exponentially increased going from brighter
to fainter magnitudes. In this way, (a) one can use a small number of stars
for the upper RGB, so that the color of the bin is not affected by the RGB slope,
and (b) it is possible to take advantage of the better statistics of the RGB
base. Finally, the brightest two stars of the RGB were not binned, and were
left as representatives of the top branch. After some experimenting, we found
that a good RGB sampling can be obtained by taking for each bin a number of
stars which is proportional to \( e^{0.2\cdot i} \), where \( i \) is an integer
number. The resulting fiducial vectors were smoothed using an average filter
with a box size of 3.

The RGB regions of the 6 clusters are shown in Fig.~\ref{f:zooms}, together
with the fiducial lines: it can be seen that in all cases the AGBs are easily
disentangled from the RGBs. The values of the fiducial points corresponding
to the solid lines in Fig.~\ref{f:zooms}, are listed in Table~\ref{t:fidtable}.

\subsection{Analytic fits to the fiducial loci\label{s:defids}}

\begin{figure}
{\par\centering \resizebox*{1\columnwidth}{!}{\includegraphics{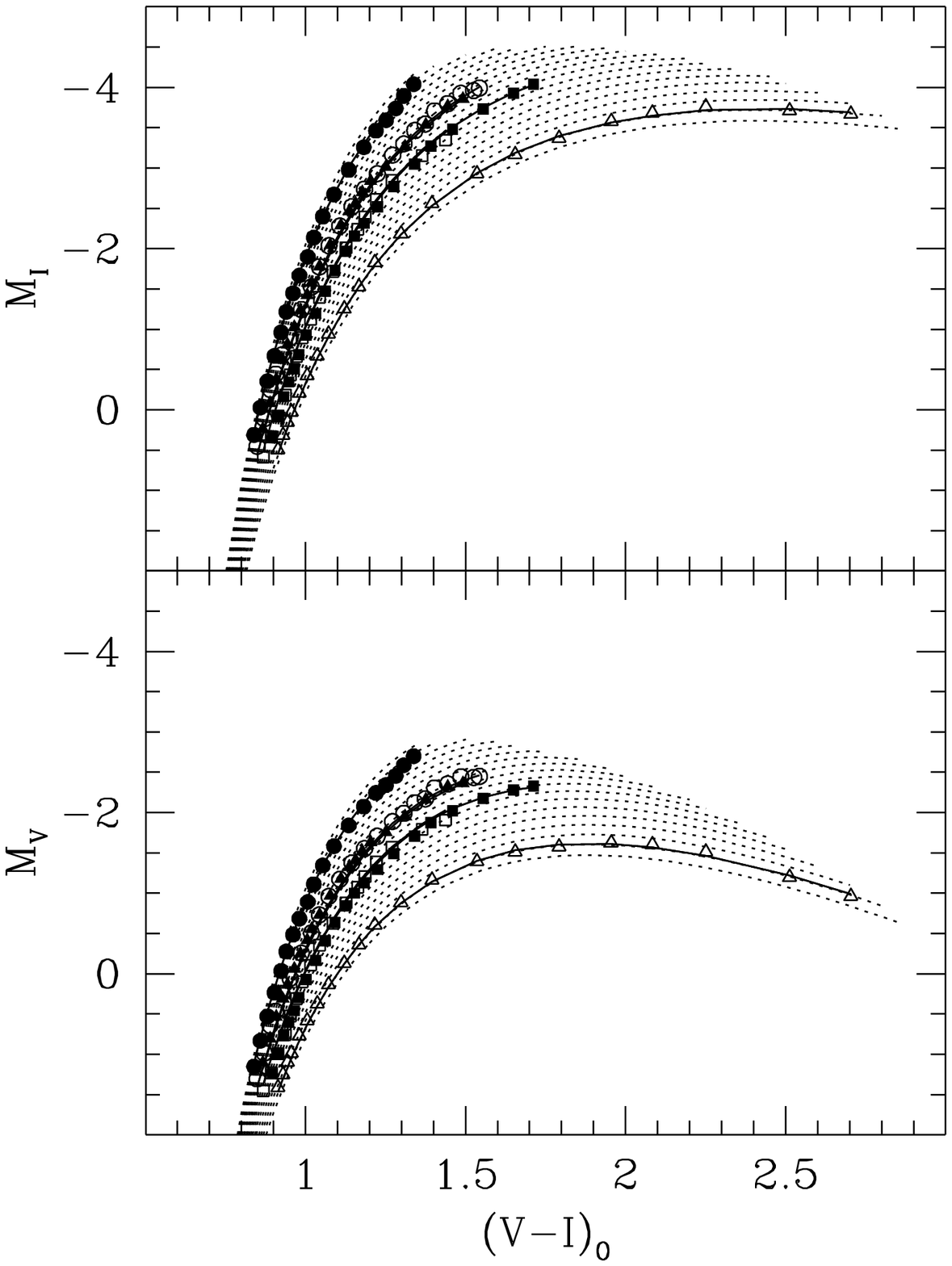}} \par}

\caption{The fiducial points of our reference sample of 6 clusters plotted over the
analytic fits for the ZW metallicity scale. The analytic RGBs (dashed
lines) have been calculated at constant $\Delta \rm [Fe/H]=0.2$~dex steps.
The observed ridge lines have been corrected for reddening and absorption +
distance scale. In the upper panel, the fits in the \protect\(
V-I,M_{I}\protect \) plane are shown, while fits in the \protect\(
V-I,M_{V}\protect \) plane are shown in the lower panel.  Different symbols
identify different clusters: NGC~104 (open triangles), NGC~288 (open
squares), NGC~5272 (open circles), NGC~5904 (solid squares), NGC~6205
(solid triangles) and NGC~6341 (solid circles) \label{f:fitsZW}}
\end{figure}

\begin{figure}
{\par\centering \resizebox*{1\columnwidth}{!}{\includegraphics{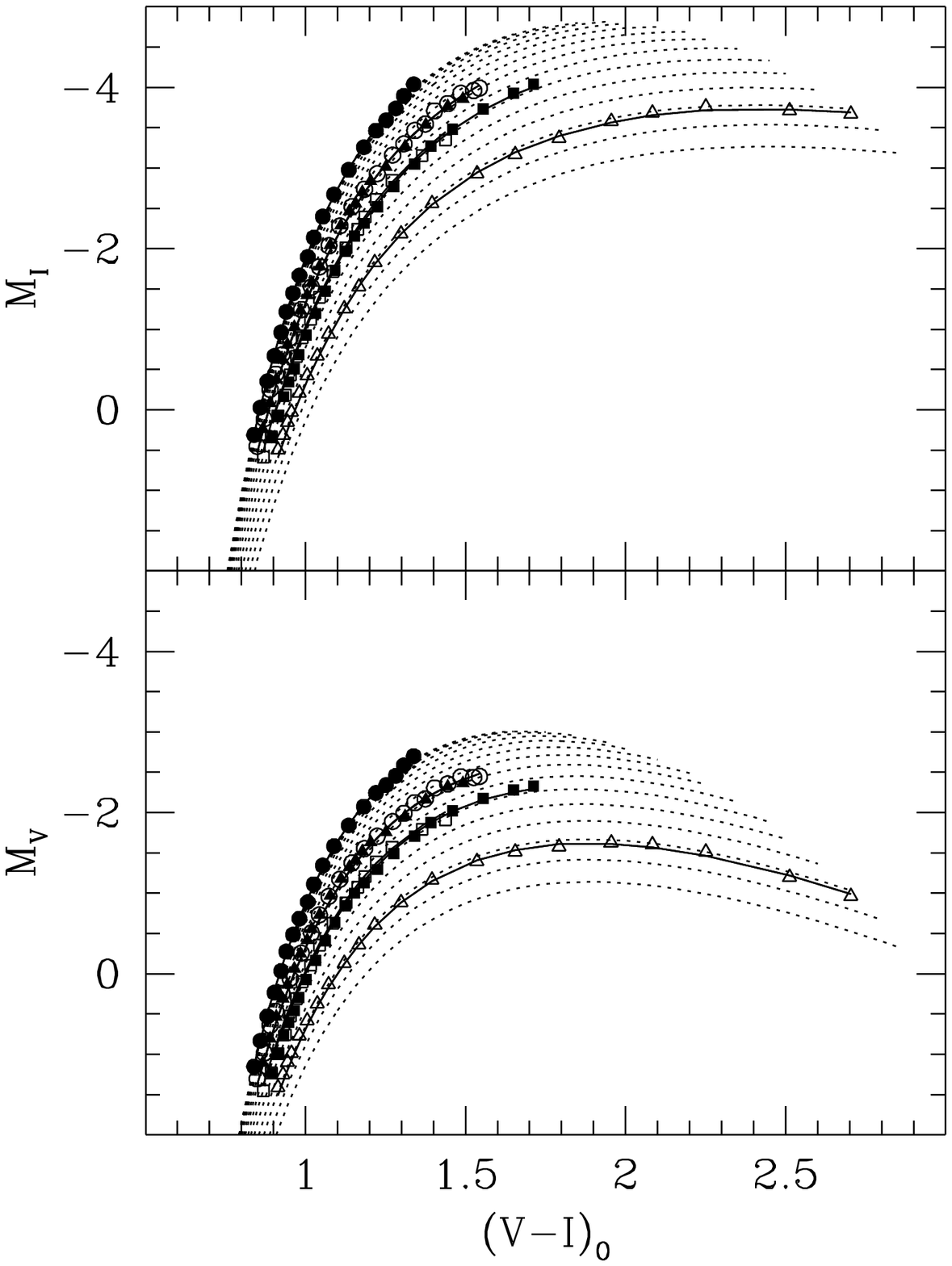}} \par}

\caption{Same as Fig.~\ref{f:fitsZW}, for the CG metallicity scale.
The metallicity step between the analytic RGBs (dashed lines) is again
$0.2$~dex. If compared to the previous figure, the non-linear trend of the
RGB color with [Fe/H] can be clearly seen
\label{f:fitsCG}}
\end{figure}
The fiducial branches defined in Sect.~\ref{s:fiducials} were fitted with a
parametrized family of hyperbolae. First, the RGBs were moved into the absolute
\( (V-I)_{0},M_{I} \) plane. The distance modulus was computed from the apparent
magnitude of the HB (cf. Table~\ref{t:fiducialGC}) and by assuming the common
law \( M_{V}({\rm HB})=a\, [{\rm Fe/H}]+b \); in order to compare our results
with those of DA90, \( a=0.17 \) and \( b=0.82 \) were used, but we also obtained
the same fits using more recent values as in Carretta et al. (\cite{cetal99}),
i.e. \( a=0.18 \) and \( b=0.90 \). The RGB was modeled with an hyperbola
as in Rosenberg et al. (\cite{rsp99}), but in this case the coefficients were
taken as second order polynomials in {[}Fe/H{]}. In other words, we parametrized
the whole family of RGBs in the following way:
\begin{equation}
\label{e:general}
M_{I}=a+b\cdot (V-I)+c/[(V-I)-d]
\end{equation}
 where 
\begin{equation}
\label{e:c123}
a=k_{1}[{\rm Fe/H}]^{2}+k_{2}[{\rm Fe/H}]+k_{3}
\end{equation}
\begin{equation}
\label{e:c456}
b=k_{4}[{\rm Fe/H}]^{2}+k_{5}[{\rm Fe/H}]+k_{6}
\end{equation}
\begin{equation}
\label{e:c789}
c=k_{7}[{\rm Fe/H}]^{2}+k_{8}[{\rm Fe/H}]+k_{9}
\end{equation}
\begin{equation}
\label{e:d}
d=k_{10}
\end{equation}
The list of the parameters of the fits in magnitude is reported in Table~\ref{t:coeffs},
together with the \emph{rms} of the residuals around the fitting curves. The
table shows that the parameter \( d \) does not depend on the choice of the
distance scale, as expected. Even the other coefficients are little dependent
on the distance scale, apart from \( k_{3} \). It is affected by the zero-point
of the HB luminosity-metallicity relation, and indeed there is the expected
\( \sim 0.1 \)~mag difference going from the LDZ to the C99 distance scale.

One could question the choice of a constant \( d \), but after some
training on the theoretical isochrones, we found that even allowing for a
varying parameter, its value indeed scattered very little around some mean
value. This empirical result is a good one, in the sense that it allows to
apply a robust linear least-square fitting method for any choice of \( d
\), and then to search for the best value of this constant by a simple
\emph{rms} minimization. We chose to fit the \( M_{I}=f\{(V-I)_{0},\rm
[Fe/H]\} \) function, and not the \( (V-I)_{0}=f(M_{I},\rm [Fe/H]) \)
function, since the latter one would be double-valued for the brightest
part of the metal rich clusters' RGBs. This choice implies that our fits
are not well-constrained for the vertical part of the giant branch,
i.e. for magnitudes fainter than \( M_{I}\sim -1 \).  However, we show in
the next section that our analytic function is good enough for the intended
purpose, i.e. to obtain the {[}Fe/H{]} of the RGB stars in far Local Group
populations, and thus to analyze how they are distributed in metallicity.

Our synthetic RGB families are plotted in Figs.~\ref{f:fitsZW} and
\ref{f:fitsCG}, for the LDZ distance scale. In the former figure, the ZW
metallicity scale is used, while the CG scale is used in the latter
one. The figures show that the chosen functional form represents a very
good approximation to the true metallicity ``distribution'' of the RG
branches. The \emph{rms} values are smaller than the typical uncertainties
in the distance moduli within the Local Group. We further stress the
excellent consistency of the empirical fiducial branches for clusters of
similar metallicity. We have two pairs of clusters whose metallicities
differ by at most 0.03~dex (depending on the scale): NGC~288 and NGC~5904
on the one side, and NGC~5272 and NGC~6205 on the other side. The figures
show that the fiducial line of NGC~288 is similar to that of NGC~5904, and
the NGC~5272 fiducial resembles that of NGC~6205, further demonstrating
both the homogeneity of our photometry and the reliability of the procedure
that is used in defining the cluster ridge lines.

If the coefficients of the hyperbolae are taken as third order polynomials,
the resulting fits are apparently better (the \emph{rms} is \( \sim 0.05 \)
mag); however, the trends of the metallicity indices show an unphysical
behavior, which is a sign that further clusters, having metallicities not
covered by the present set, would be needed in order to robustly constrain
the analytic function.

In the following section, the indices are calibrated in terms of
metallicity, so that in Sect.~\ref{s:testfits} they will be used to check
the reliability of our generalized fits.

\section{Calibration of the indices. Introduction}

\begin{table}

\caption{The coefficients that define the functions used to interpolate our RGBs (see
text); the top header line identifies the two distance scales used, while the
two metallicities are identified in the second line of the header\label{t:coeffs}}

\begin{tabular}{rrrrr}
\hline\hline
\noalign{\smallskip}&
\multicolumn{2}{c}{LDZ}&
\multicolumn{2}{c}{C99}\\
&
 CG &
 ZW &
 CG &
 ZW \\
\hline 
\noalign{\smallskip} 
  d &
 0.212 &
 0.182 &
0.212 &
 0.182\\
k1 &
 -0.231 &
 -1.338 &
-0.227 &
-1.336\\
 k2 &
 3.290 &
 -0.069 &
3.314 &
-0.055 \\
 k3 &
 -7.229 &
 -9.547 &
-7.140 &
-9.465\\
 k4 &
 0.611 &
 0.710 &
0.612 &
0.709\\
 k5 &
 0.551 &
 0.883 &
0.556 &
0.881\\
 k6 &
 1.398 &
 1.651 &
1.401 &
1.650 \\
 k7 &
 0.380 &
 0.525 &
0.381 &
0.524\\
 k8 &
 -0.135 &
 0.206 &
-0.133 &
0.204\\
 k9 &
 6.194 &
 6.806 &
6.195 &
6.805 \\
 {\em rms} &
 0.07 &
 0.08 &
 0.07 &
 0.08  \\
\hline 
\end{tabular}

\end{table}
\begin{table}

\caption{Coefficients of the calibrating relations for the indices (see text for the
definition of the equations). NGC~6656 was excluded from the fits \label{t:rms}}

\begin{tabular}{cccccccc}
\hline
\hline\noalign{\smallskip}
{\scriptsize index}&
{\scriptsize d.sc.}&
{\scriptsize metallicity}&
{\scriptsize \( \alpha \)}&
{\scriptsize \( \beta \)}&
{\scriptsize \( \gamma \)}&
\emph{\scriptsize rms}{\scriptsize }&
{\scriptsize fit}\\
\noalign{\smallskip}
\hline\noalign{\smallskip}
{\scriptsize \( S \)}&
{\scriptsize }&
{\scriptsize CG}&
{\scriptsize -0.03}&
{\scriptsize 0.23}&
{\scriptsize -1.19}&
{\scriptsize 0.13}&
{\scriptsize 2}\\
{\scriptsize }&
{\scriptsize }&
{\scriptsize ZW}&
{\scriptsize -0.004}&
{\scriptsize -0.18}&
{\scriptsize 0.08}&
{\scriptsize 0.12}&
{\scriptsize 2}\\
{\scriptsize }&
{\scriptsize }&
{\scriptsize ZW}&
{\scriptsize -0.24}&
{\scriptsize 0.28}&
{\scriptsize }&
{\scriptsize 0.12}&
{\scriptsize 1}\\
{\scriptsize \( (V-I)_{-3.5} \)}&
{\scriptsize LDZ}&
{\scriptsize CG }&
{\scriptsize 0.00487}&
{\scriptsize -0.0057}&
{\scriptsize }&
{\scriptsize 0.13}&
{\scriptsize \( z \) }\\
{\scriptsize }&
{\scriptsize }&
{\scriptsize ZW }&
{\scriptsize -2.12}&
{\scriptsize 8.81}&
{\scriptsize -9.75}&
{\scriptsize 0.13}&
{\scriptsize 2}\\
{\scriptsize }&
{\scriptsize C99}&
{\scriptsize CG {}{}}&
{\scriptsize 0.0045}&
{\scriptsize -0.0053}&
{\scriptsize }&
{\scriptsize 0.15}&
{\scriptsize \( z \) }\\
{\scriptsize }&
{\scriptsize }&
{\scriptsize ZW {}{}}&
{\scriptsize -2.05 }&
{\scriptsize 8.57 }&
{\scriptsize -9.61 }&
{\scriptsize 0.12}&
{\scriptsize 2}\\
{\scriptsize \( (V-I)_{-3.0} \)}&
{\scriptsize LDZ}&
{\scriptsize CG}&
{\scriptsize 0.0068}&
{\scriptsize -0.0076}&
{\scriptsize }&
{\scriptsize 0.15}&
{\scriptsize \( z \)}\\
{\scriptsize }&
{\scriptsize }&
 {\scriptsize ZW}&
{\scriptsize -3.34}&
{\scriptsize 12.37}&
{\scriptsize -11.91}&
{\scriptsize 0.14}&
{\scriptsize 2 }\\
{\scriptsize }&
{\scriptsize C99}&
{\scriptsize CG {}{}}&
{\scriptsize 0.0065}&
{\scriptsize -0.0073}&
{\scriptsize }&
{\scriptsize 0.15}&
{\scriptsize \( z \)}\\
{\scriptsize }&
{\scriptsize }&
{\scriptsize ZW {}{}}&
{\scriptsize -3.233 }&
{\scriptsize 12.23 }&
{\scriptsize -11.96 }&
{\scriptsize 0.14}&
{\scriptsize 2 }\\
{\scriptsize \( \Delta V_{1.4} \)}&
{\scriptsize }&
{\scriptsize CG}&
{\scriptsize -0.34}&
{\scriptsize 0.93}&
{\scriptsize -1.37}&
{\scriptsize 0.16}&
{\scriptsize 2}\\
{\scriptsize }&
{\scriptsize }&
{\scriptsize ZW}&
{\scriptsize -0.063}&
{\scriptsize -0.56}&
{\scriptsize 0.41}&
{\scriptsize 0.16}&
{\scriptsize 2}\\
{\scriptsize }&
{\scriptsize }&
{\scriptsize ZW}&
{\scriptsize -0.87}&
{\scriptsize 0.77}&
{\scriptsize }&
{\scriptsize 0.16}&
{\scriptsize 1}\\
{\scriptsize \( \Delta V_{1.2} \)}&
{\scriptsize }&
{\scriptsize CG}&
{\scriptsize -0.36}&
{\scriptsize 0.55}&
{\scriptsize -0.97}&
{\scriptsize 0.19}&
{\scriptsize 2}\\
{\scriptsize }&
{\scriptsize }&
{\scriptsize CG}&
{\scriptsize -0.69}&
{\scriptsize 0.0007}&
{\scriptsize }&
{\scriptsize 0.22}&
{\scriptsize 1}\\
{\scriptsize }&
{\scriptsize }&
{\scriptsize ZW}&
{\scriptsize -0.13}&
{\scriptsize -0.38}&
{\scriptsize -0.28}&
{\scriptsize 0.20}&
{\scriptsize 2}\\
{\scriptsize }&
{\scriptsize }&
{\scriptsize ZW}&
{\scriptsize -0.82}&
{\scriptsize 0.06}&
{\scriptsize }&
{\scriptsize 0.20}&
{\scriptsize 1}\\
{\scriptsize \( \Delta V_{1.1} \)}&
{\scriptsize }&
{\scriptsize CG}&
{\scriptsize -0.30}&
{\scriptsize 0.09}&
{\scriptsize -0.81}&
{\scriptsize 0.23}&
{\scriptsize 2}\\
{\scriptsize }&
{\scriptsize }&
{\scriptsize CG}&
{\scriptsize -0.59}&
{\scriptsize -0.52}&
{\scriptsize }&
{\scriptsize 0.25}&
{\scriptsize 1}\\
{\scriptsize }&
{\scriptsize }&
{\scriptsize ZW}&
{\scriptsize -0.13}&
{\scriptsize -0.42}&
{\scriptsize -0.68}&
{\scriptsize 0.25}&
{\scriptsize 2}\\
{\scriptsize }&
{\scriptsize }&
{\scriptsize ZW}&
{\scriptsize -0.70}&
{\scriptsize -0.56}&
{\scriptsize }&
{\scriptsize 0.25}&
{\scriptsize 1}\\
{\scriptsize \( (V-I)_{0,\rm g} \)}&
{\scriptsize }&
{\scriptsize CG}&
{\scriptsize 4.25}&
{\scriptsize -5.37}&
{\scriptsize }&
{\scriptsize 0.32}&
{\scriptsize 1}\\
{\scriptsize }&
{\scriptsize }&
{\scriptsize ZW}&
{\scriptsize 5.25}&
{\scriptsize -6.52}&
{\scriptsize }&
{\scriptsize 0.33}&
{\scriptsize 1}\\
\noalign{\smallskip}
\hline
\end{tabular}

\end{table}
\begin{figure}
{\par\centering \resizebox*{1\columnwidth}{!}{\includegraphics{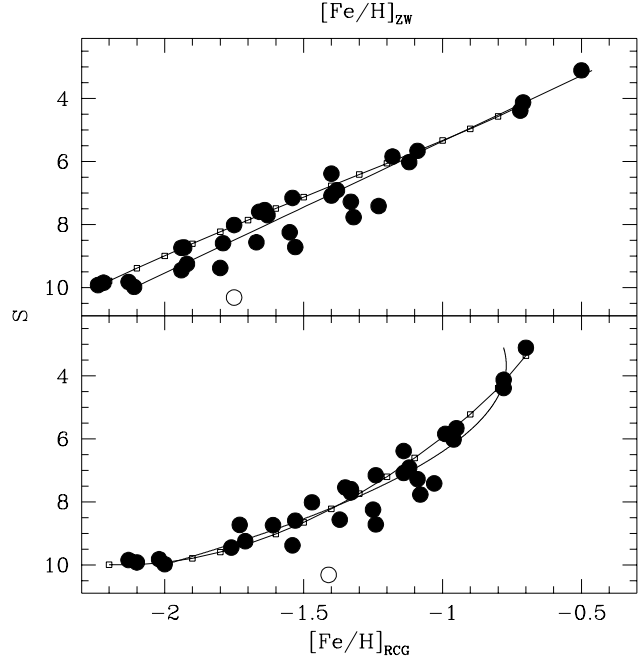}} \par}

\caption{Calibration of the index \protect\( S\protect \) (cf. Fig.~\ref{f:pars-parteA},
left panel) as a function of \protect\( [\rm Fe/H]\protect \) on the Zinn \&
West (1984) scale (\emph{top panel}) and on the Carretta \& Gratton (1997) scale
(\emph{bottom panel}). Linear (top panel) and parabolic (bottom panel) fits
of the data are also represented. The cluster marked with open circle was excluded
from the fit (see text for details). Starting from this figure (to Fig.~\ref{f:calDV14}),
the open squares connected by a solid line represent the mono-parametric approximation
(see Sect.~\ref{s:newda90}) \label{f:calS}}
\end{figure}
\begin{figure}
{\par\centering \resizebox*{1\columnwidth}{!}{\includegraphics{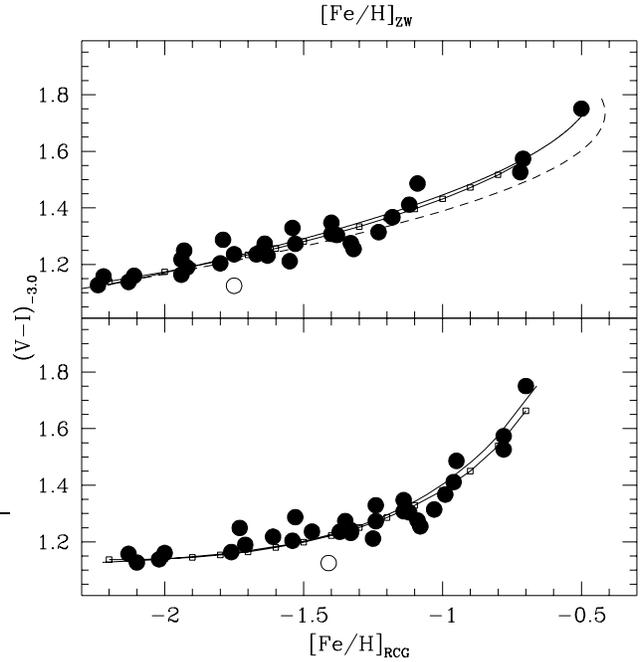}} \par}

\caption{Calibration of the index \protect\( (V-I)_{-3.0}\protect \). The solid lines
represent the equations described in the text, while the dashed curve represents
the the DA90 calibration. \label{f:calVI-3}}
\end{figure}

\begin{figure}
{\par\centering \resizebox*{1\columnwidth}{!}{\includegraphics{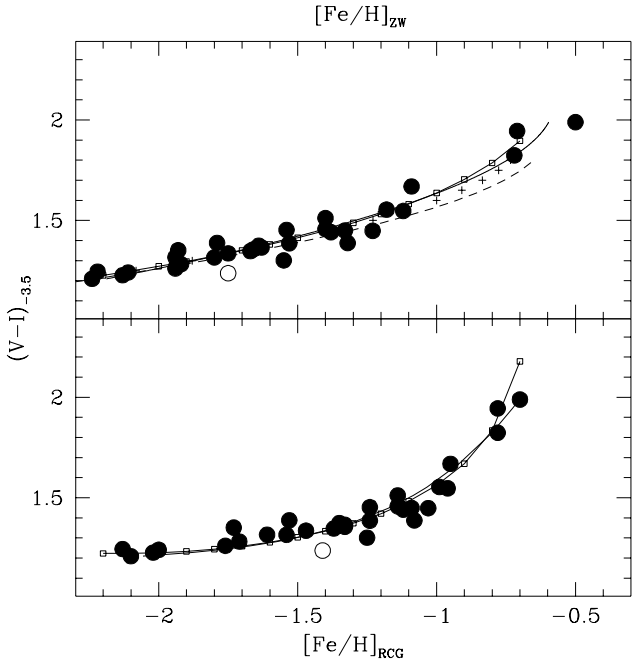}} \par}

\caption{Calibration of the index \protect\( (V-I)_{-3.5}\protect \) The solid lines
represent the equations described in the text, while the dashed curve
represents the the Lee et al. (1993) calibration. 
The Caldwell et
al. \cite{caldwell-3p5} relation is also shown with plus symbols
\label{f:calVI-3.5}}
\end{figure}

\begin{figure}
{\par\centering \resizebox*{1\columnwidth}{!}{\includegraphics{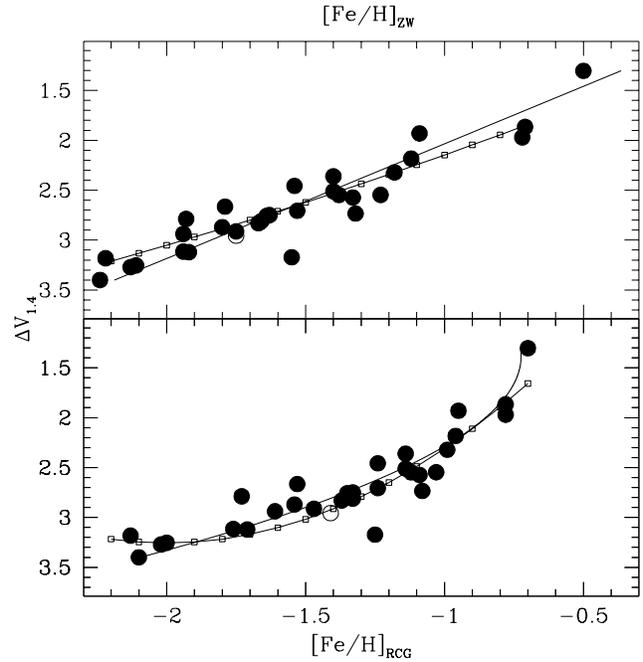}} \par}

\caption{Calibration of the index \protect\( \Delta V_{1.4}\protect \) \label{f:calDV14}.
The solid lines represent linear (top panel) and quadratic (bottom panel) fits
to the data}
\end{figure}

In order to obtain analytic relations between the indices and the actual
metallicity, our photometric parameters were compared both with the ZW and
the CG values.  A summary of the resulting equations is given in
Table~\ref{t:rms}. For each index (first column) both linear and quadratic
fits were tried, of the form:
\( \rm [Fe/H]=\alpha \cdot index +\beta  \) and \( \rm [Fe/H]=\alpha
\cdot index^{2}+\beta \cdot index +\gamma  \).
The coefficients of the calibrating relation are given in the columns
labelled
\( \alpha  \), \( \beta  \), and \( \gamma  \); in column 7, the \emph{rms}
of the residuals is also given. In the case of the \( (V-I)_{-3.0} \) and
\( (V-I)_{-3.5} \) indices, neither the linear nor the quadratic fits give
satisfactory results, when the CG scale is considered. Instead, a good fit
is obtained if a change of variables is performed, setting \( z=0.02\times
10^{[{\rm Fe/H}]} \), and linearly interpolating in the index (i.e. setting
\(  z=\alpha \cdot \rm index +\beta \)).  The column 8 of Table~\ref{t:rms}
identifies the kind of fitting function that is used for each
parameter/metallicity combination: the symbols ``1'', ``2'' and ``\( z \)''
refer to the linear, quadratic, and linear in \( z \) fits,
respectively. Relations on both the CG and ZW metallicity scales are given,
and column 3 flags the {[}Fe/H{]} scale that is used.

In order to measure the \( (V-I)_{-3} \) and \( (V-I)_{-3.5} \) indices
(cf.  Sect.~\ref{s:indices}) a distance scale must be adopted. The most
straightforward way is to use the observed \( V_{\rm HB} \)
(cf. Table~\ref{t:the-sample}) coupled with a suitable law for the HB
absolute magnitude.

It has become customary to parameterize this magnitude as \( M_{V}({\rm
HB})=a\cdot [{\rm Fe/H}]+b \), although there is no consensus on the value
of the two parameters \( a \) and
\( b \). The current calibrations of these two metallicity indices were obtained
by Da Costa \& Armandroff (\cite{da90}) and Lee et al. (\cite{lee.et.al93}),
and they are based on the Lee et al. (\cite{ldz}; LDZ) theoretical luminosities
of the HB. LDZ gave a relation \( M_{V}({\rm HB})=0.17\cdot [{\rm Fe/H}]+0.82 \)
valid for \( Y=0.23 \). 

As discussed in Sect~\ref{s:newda90}, since many current determinations of
Population~\noun{ii} distances within the Local Group are based on the Lee
et al. (\cite{ldz}) distance scale, and for the purpose of comparison with
previous studies, we provide a calibration using the latter HB
luminosity-metallicity relation. However, in the last ten years revisions
of this relation have been discussed by many authors, so we also calibrated
the two indices using \( M_{V}({\rm HB})=0.18\cdot [{\rm Fe/H}]+0.90 \)
(Carretta et al. \cite{cetal99}), which is one of the most recent HB-based
distance scales.

We must stress that \emph{metallicities on the ZW scale must be used in the
\( M_{V} \) vs.} {[}Fe/H{]} \emph{relation}. Indeed, CG showed that their scale
is not linearly correlated to that of ZW, so not even the \emph{\( M_{V} \)
vs.} {[}Fe/H{]} relation will be linear: if one wishes to use the new scale,
then \emph{the absolute magnitude of the HB must be re-calibrated} in a more
complicated way. 

The best calibrating relations are shown in Figs.~\ref{f:calS} to
\ref{f:calDV14}.  In the following sections, for each index a few remarks
on the accuracy of the calibrations and comparisons with past studies are
given.

\section{Calibration of the indices. Discussion\label{s:calibrations}}

\subsection{\protect\( S\protect \)}

On the CG scale, the second-order fit has a residual \emph{rms} of 0.12~dex
in {[}Fe/H{]}. On the ZW scale, the linear fit is obtained with a
\emph{rms} of 0.12~dex. This index can therefore be calibrated on both
scales, with a comparable level of accuracy. A parabolic fit does not
improve the relation on the ZW scale, since the coefficient of the
quadratic term is very small (-0.004) and the \emph{rms} is the same. These
relations are shown in Fig.~\ref{f:calS} as solid lines, where the upper
panel is for the ZW scale, and the lower panel for the CG scale (this
layout is reproduced in all the following figures).

The cluster NGC~6656 (M22) was excluded from the fits, and is plotted as an
open circle in Fig.~\ref{f:calS}. It is well-known that M22 is a cluster that
shows a metallicity spread, and indeed it falls outside the general trend in
most of the present calibrations.

\subsection{\protect\( (V-I)_{-3.0}\protect \) \label{s:DA90indices1}}

The first definition of the \( (V-I)_{-3.0} \) index was given in Da Costa
\& Armandroff (\cite{da90}), where a calibration in terms of the ZW scale was
also given: \( [{\rm Fe/H}]=-15.16+17.0\, (V-I)_{-3}-4.9\, (V-I)_{-3}^{2} \).
The same index (measured on the \emph{absolute} RGBs corrected with the LDZ
HB luminosity-metallicity relation) is plotted, in Fig.~\ref{f:calVI-3}, as
a function of the metallicity on both scales, and the solid lines represent
our calibrations. The top panel shows the quadratic relation on the ZW scale,
whose \emph{rms} is 0.14~dex. The bottom panel of Fig.~\ref{f:calVI-3} shows
the relation on the CG scale. In this case, a quadratic fit is not able to reproduce
the trend of the observational data. A better result can be obtained by making
a variable change, i.e. using the variable \( z=0.02 \cdot 10^{[{\rm Fe/H}]} \);
in this case, a linear relation is found, and its \emph{rms} is 0.15~dex. This
measure of the residual scatter has been computed after transforming back to
metallicity, so the reliability of the index can be compared to that of the
other ones. Again, the index can be calibrated on both scales with a comparable
accuracy. The dashed curve in the upper panel of Fig.~\ref{f:calVI-3} shows
the original relation obtained by DA90: there is a small discrepancy at the
high-metallicity end, which can be explained by the different 47~Tuc fiducial
line that was adopted by DA90 (cf. below the discussion on \( (V-I)_{-3.5} \)). 

As already recalled, we checked the effect of adopting another distance scale,
by repeating our measurements and fits, and adopting the C99 distance scale.
For the ZW metallicity scale, we obtain the quadratic relation whose coefficients
are listed in Table~\ref{t:rms}, and whose \emph{rms} is 0.15~dex. The bottom
panel of Fig.~\ref{f:calVI-3} shows the relation on the CG scale. Again, a
quadratic fit is not able to reproduce the trend of the observational data.
Making the already discussed variable substitution, the linear relation in \( z \)
has an \emph{rms} of 0.16~dex, so the two metallicity scales yield almost comparable
results.

\subsection{\protect\( (V-I)_{-3.5}\protect \)\label{s:DA90indices2}}

Using the same ``standard'' GC branches of DA90, Lee et
al. (\cite{lee.et.al93}) defined a new index, \( (V-I)_{-3.5} \), to be
used for the farthest population~\noun{ii} objects. It was also calibrated
in terms of the ZW scale: \( [{\rm Fe/H}]=-12.64+12.6\, (V-I)_{-3.5}-3.3\,
(V-I)_{-3.5}^{2} \).  
A new calibration was also given recently in Caldwell et
al. (\cite{caldwell-3p5}): [Fe/H]$=-1.00 + 1.97\,q - 3.20\,q^2$, where
$q=[(V-I)_{-3.5}-1.6]$.
The index and our calibrations (solid lines) are
plotted, in Fig.~\ref{f:calVI-3.5}, on both metallicity scales. Again, the
measurements were made in the absolute CMD, assuming the LDZ distance
scale. Our quadratic calibration vs. the ZW scale has a residual \emph{rms}
scatter of 0.13~dex, which is the same of the linear relation on the CG
metallicity vs. \( z \).

The Lee et al. relation (dashed
line) predicts slightly too larger metallicities on the ZW scale, for \( \rm [Fe/H]>-1 \).
This can also be interpreted as if the DA90 47~Tuc branch were \( <0.1 \) mag
bluer than ours. Indeed, if one looks at Fig.~5 of DA90, one can easily see
that some weight is given to the brightest RGB star, which is brighter than
the trend defined by the previous ones. The result is a steeper branch, which
also justifies the DA90 slightly bluer RGB fiducial. Since our metal richest
point is defined by two clusters, and since the two measured parameters agree
very well, we are confident that our calibration is reliable. In any case, the
discrepancy between the two scales is no larger than \( \sim 0.1 \)~dex.
It is also reassuring that the Caldwell et al. (\cite{caldwell-3p5})
relation (pluses) is closer to the present calibration, since the former
is based on a larger set of clusters. This might be an indication that the
Lee et al. relation is actually inaccurate at the metal rich end, due to
the small set of calibrating clusters.

As before, we obtained a further calibration also using the C99 \( M_{V} \)
vs. {[}Fe/H{]} relation; the quadratic fit on the ZW scale has a residual \emph{rms}
scatter of 0.13~dex, while the \( z \) variable can be fitted with a straight
line, with an \emph{rms} of 0.14~dex.

\subsection{The \protect\( \Delta V\protect \) family and \( (V-I)_{\rm
0,g}\) }

For any \( \Delta V \) index, the quadratic relations vs. the ZW
metallicity do not improve the \emph{rms} and they are not plotted in the
figures. The coefficients are listed in Table~\ref{t:rms}.

The best metallicity estimates of the ``\( \Delta V \) family'' are obtained
with the \( \Delta V_{1.4} \) index. The errors on \( [{\rm Fe/H}] \)
are just slightly larger than the standard uncertainties of the spectroscopic
determinations. The solid lines of Fig.~\ref{f:calDV14} show the calibrations
that we obtain. The quadratic equation on the CG scale, and the linear one on
the ZW scale, are obtained with residual scatters of 0.16~dex.

The rest of the indices in this family, and \( (V-I)_{\rm 0,g}\), lack the
precision of the other abundance indicators. This is due to the fact that
the error on any \( \Delta V \) index is proportional to the uncertainty on
the color of the RGB (which depends on the reddening), times its local
slope where the reference point is measured. Since the RGB slope increases
going away from the tip (i.e. towards bluer colors), we expect that the
scatter on the \( \Delta V \) indices will also increase as the color of
the reference point gets bluer. Indeed, Table~\ref{t:rms} shows that in
most cases the {\em rms} uncertainties are $> 0.2$~dex for these
indices. The residual scatter is largest for the \( (V-I)_{\rm 0,g}\)
index, which is the most affected by the uncertainties on the reddening.

The \( \Delta V_{1.2} \) and \( (V-I)_{\rm 0,g}\) parameters have been
earlier calibrated, on the CG scale, by Carretta \& Bragaglia
(\cite{cb98}). Using their quadratic relation for \( \Delta V_{1.2} \), and
both their linear and quadratic relations for \( (V-I)_{\rm 0,g}\), the
corresponding \emph{rms} of the residuals in metallicity are 0.21~dex and
\( \sim 0.41 \)~dex, respectively. 
 Our new and the old calibrations are therefore compatible, within the
 (albeit large) uncertainties.

\section{A test of the ``model'' RGBs; comparison with the observed {[}Fe/H{]} indices\label{s:testfits}}

A straightforward test of our new analytic RGBs can be made by generating
the same metallicity indices that have been measured on the observed RGBs,
and then checking the consistency of the predicted vs. measured
quantities. To this aim, for a set of discrete {[}Fe/H{]} values a \(
(V-I)_{0} \) vector was generated, and the combination of the two was used
to compute the \( M_{I} \) vector of the giant branch, using
Eqs. (\ref{e:general}-\ref{e:d}). Then for each branch the metallicity
indices were measured as it was done for the clusters' fiducials.  

In Figs.~\ref{f:calS} to \ref{f:calDV14}, the predicted indices are
identified by the small open squares (spaced by 0.1~dex) connected by a
solid line.  
The best predictions are for those indices that rely on the brightest part
of the RGB (i.e. \( (V-I)_{-3.0} \), \( (V-I)_{-3.5} \) and \( \Delta
V_{1.4} \)), while the computations are partially discrepant for those
indices that rely on a point that is measured on the faint RGB. This is
easily explained by the nature of our fit: since the best match is searched
for along the ordinates (for the reasons discussed in
Sect.~\ref{s:newda90}), then it is better constrained in the upper part of
the RGB, where its curvature becomes more sensitive to metallicity. We must
also stress that the metal richest cluster in the reference grid is 47~Tuc
({[}Fe/H{]}\( =-0.70 \) on the ZW scale), whereas NGC~6352 ({[}Fe/H{]}\(
=-0.50 \) on the same scale) is the metal richest cluster for which
metallicity indices have been measured. Some of the discrepancies that are
seen at the highest metallicities are therefore due to the lack of
low-reddening clusters that can be used to extend the reference grid to the
larger {[}Fe/H{]} values.

The mean differences between the predicted and fitted indices are, on the ZW
scale, around 0.03~dex for the \( (V-I)_{-3.0} \) and \( (V-I)_{-3.5} \) indices.
They are around 0.08~dex for the \( \Delta V_{1.2} \), \( \Delta V_{1.4} \),
and \( S \) indices. They rise to \( \sim 0.1 \) and \( \sim 0.3 \)~dex for
the \( \Delta V_{1.1} \) and \( (V-I)_{0,\rm g} \) indices. A similar trend
is seen for the comparison on the CG scale. In this case, the mean differences
are \( \sim 0.05 \)~dex for \( (V-I)_{-3.0} \), \( (V-I)_{-3.5} \), and \( S \);
they are \( \sim 0.1 \)~dex for \( \Delta V_{1.2} \) and \( \Delta V_{1.4} \);
and they are 0.12 and 0.27 for the \( \Delta V_{1.1} \) and \( (V-I)_{0,\rm g} \)
indices.

We can therefore conclude that, apart from the \( \Delta V_{1.1} \) and \( (V-I)_{0,\rm g} \)
indices, our mono-parametric RGB family gives a satisfactory reproduction of
the actual changes of the RGB morphology and location, as a function of metallicity.
It is then expected that, using this approach, one can exploit the brightest
\( \sim 3 \)~mags of the RGB to determine the mean metallicity, and even more
important, the metallicity \emph{distribution} of the old stellar population
of any Local Group galaxy. In a forthcoming paper, we will demonstrate such
possibility by re-analyzing our old photometric studies of the dwarf spheroidal
galaxies Tucana (Saviane et al. \cite{tucana}), Phoenix (Held et al. \cite{phoenix};
Mart\'{\i}nez-Delgado et al. \cite{aaj_phoenix}), Fornax (Saviane et al. \cite{fornax}),
LGS~3 (Aparicio et al. \cite{aaj_lgs3}), Leo~I (Gallart et
al. \cite{aaj_leoi}; Held et al. \cite{held_leoi})
and NGC~185 (Mart\'{\i}nez-Delgado et al. \cite{aaj_185}).

\section{Conclusions \label{s:conclusioni}}

In this work, we have provided the first calibration of a few metallicity indices
in the \( (V-I),V \) plane, namely the indices \( S \), \( \Delta V_{1.1} \)
and \( \Delta V_{1.4} \). Calibrations on both the Zinn \& West (1984) and
Carretta \& Gratton (1997) scales have been obtained. The metallicity indices
\( (V-I)_{0,\rm g} \), \( \Delta V_{1.2} \) , \( (V-I)_{-3.0} \) and \( (V-I)_{-3.5} \)
have been also calibrated on both scales, and we have shown that our new relations
are consistent with existing ones. In the case of the latter two indices, we
have obtained the first calibration on the CG scale; for both scales, we have
also obtained the first calibration that takes into account new results on the
RR~Lyr distances . The accuracy of the calibrations is generally better than
0.2~dex, regardless of the metallicity scale that is used.

Our results are an improvement over previous calibrations, since a new approach
in the definition of the RGB is used, and since our formulae are based on the
largest homogeneous photometric database of Galactic globular clusters. 

The availability of such database also allowed us a progress towards the definition
of a standard description of the RGB morphology and location. We were able to
obtain a function in the \( (V-I)_{0},M_{I},\rm [Fe/H] \) space which is able
to reproduce the whole set of GGC giant branches in terms of a single parameter
(the metallicity). We suggest that the usage of this function will improve the
current determinations of metallicity and distances within the Local Group,
extending the methods of Lee et al. (1993).

\begin{acknowledgements}
We thank the referee, Gary Da Costa, for helpful suggestions that improved
the final presentation of the manuscript. I.S. acknowledges the financial
support of Italian and Spanish Foreign Ministries, through an `Azioni
Integrate/Acciones Integradas' grant.
\end{acknowledgements}

\end{document}